\documentstyle[preprint,aps]{revtex}
\begin{document}
\draft
\preprint{UPR-717-T}
\date{September 1996}
\title{Supersymmetric dyonic black holes of IIA string on Six Torus}
\author{Kwan-Leung Chan
\thanks{E-mail address: klchan@cvetic.hep.upenn.edu}}
\address{Department of Physics and Astronomy \\
         University of Pennsylvania, Philadelphia PA 19104-6396}
\maketitle
\begin{abstract}
{A class of four-dimensional static supersymmetric black hole solutions of effective supergravity Lagrangian of IIA superstring compactified on $T^6$ is constructed by explicitly solving Killing spinor equations (KSEs). These solutions are dyonic black holes parametrized by four charges, with dilaton and diagonal internal metric components as the only non-zero scalar fields, and preserve $1 \over 8$ of $N=8$ supersymmetry. The KSEs with only Neveu-Schwarz-Neveu-Schwarz charges relate spinors with opposite chirality from ten-dimensional view point, and have identical structures with KSEs of toroidally compactified heterotic string. We also find a solution with four Ramond-Ramond charges which is U-dual to the solution with four Neveu-Schwarz-Neveu-Schwarz charges, and corresponds to the intersecting D-brane configuration with two 2-branes and two 4-branes. A configuration with both Neveu-Schwarz-Neveu-Schwarz charges and Ramond-Ramond charges is also found. We show that the configurations T-dual to the above solutions are also solutions of the KSEs. The patterns of supersymmetry breaking are studied in detail.}
\end{abstract}

\section{Introduction and Summary}

Supersymmetric black hole solutions with masses saturating the corresponding Bogomol'nyi bounds, ${\it i.e.}$, BPS-saturated solutions of superstring theories, have been a subject of much research recently. Being non-perturbative in nature, it is important in understanding non-perturbative duality symmetries in string theory. In view of the non-perturbative nature of the proposed duality conjecturs [1]-[12], which led to the unifying treatment of all five superstring theories ( Type I, Type IIA, Type IIB, heterotic SO(32), heterotic $E_8 \times E_8$), ${\it i.e.}$,  each corresponds to a corner of the same underlying M-theory[13], the role of BPS-saturated solutions becomes crucial. Such BPS-saturated states can become massless in special region of moduli space which parametrizes the underlying string vacua [14]-[20], and thereby enhance gauge symmetry as well as supersymmetry. Another good reason for the study comes from the recent dramatic progress that has been made in the understanding of the microscopic origin of the black hole entropy [21]. The degeneracy of BPS-saturated states is shown to have clear connection with the statistical nature of the internal structures of black holes.

The program of finding general spherically symmetric, static four-dimensional BPS-saturated black hole solutions of effective supergravity Lagrangian of heterotic string compactified on six torus has been completed [22]-[25]. In [22], the Killing spinor equations (KSEs) were solved, and a generating solution parametrized by four independent charges was constructed. A more general solution was obtained by performing S-duality and T-duality transformations on the generating solution. They were parametrized by 55 independent charges, ${\it i.e.}$, 56 charges with one charge constraint. In [23], the generating solution was shown to be an exact solution (to all orders in world sheet expansion) of a string theory. It was further generalized in [24] to a generating solution with five independent charges, all of which were defined in two toroidally compactified directions. By performing duality transformations on this generating solution, one found the most general solution which was parametrized by 56 independent charges. 

The purpose of this paper is to find static, spherically symmetric BPS-saturated black hole solutions of low energy effective IIA superstring theory on six-torus, by explicitly solving KSEs. We turn off all the scalar fields except dilaton and components of the diagonal internal metric. In [26], static, spherically symmetric BPS-saturated states parametrized by two charges were constructed. They correspond to the special cases when either the Kaluza-Klein fields or the three-form fields of the underlying N=1 11-dimensional supergravity are turned on, but not both. Here we find the more general configurations with both fields turned on. With this more general setting, we can explicitly show the embedding of the $N=4$ supersymmetry of the toroidally compactified heterotic string within the $N=8$ supersymmetry of the toroidally compactified IIA superstring (when only Neveu-Schwarz-Neveu-Schwarz charges are non-zero). Moreover, we can find the BPS-saturated solutions with four Ramond-Ramond charges and those with both Ramond-Ramond charges and Neveu-Schwarz-Neveu-Schwarz charges. We can also study the pattern of supersymmetry breaking explicitly.

We start with the $N=1$ 11-dimensional supergravity (SG) theory in Section II. We compactify this supergravity theory on a seven-torus, $T^7$ [27][28], and obtain the corresponding effective four-dimensional action. Then we express the same four-dimensional action in terms of fields of IIA superstring compactified on a six-torus, $T^6$. The four-dimensional fields of the toroidally compacified IIA superstring are obtained in two steps. We first compactify the 11-d SG on a circle, $S^1$, and identify the resulting 10-d fields with the fields of IIA superstring in 10-d. Then we compactify the 10-d action of IIA superstring on $T^6$. The field redefinition rules that relate the four-dimensional fields from SG on $T^7$ and that from IIA on $T^6$ are needed to get the KSEs of IIA superstring from the KSEs of SG, which are obtained straightforwardly by compactifying the transformation rule of the gravitino of SG in 11 dimensions down to four dimensions on a seven torus. 

We obtain the KSEs of the IIA superstring compactified on $T^6$ in Section III. We express the supersymmetry transformation rules of the gravitinos and modulinos of the compactified SG in terms of four-dimensional fields, which are obtained by compactifying the 11-dimensional SG on $T^7$ directly, at the beginning. Then in Section IIIA, we simplify the KSEs by turning off all scalar fields except the dilaton and the diagonal internal metric elements. We also assume spherical symmetry and time-independence. With the field redefinition rules derived in Section II, we find the KSEs of the toroidally compactified IIA superstring from the KSEs of the compactified SG in Section IIIB. These KSEs involve the charges from the Neveu-Schwarz-Neveu-Schwarz (NS-NS) sector and the Ramond-Ramond (RR) sector. There are two types of spinors originating from the two 10-dimensional spinors with opposite chiralities associated with the $N=2$ supersymmetry of the IIA superstring in 10 dimensions. In Section IIIC, we study the spinor constraints, which determine the patterns of supersymmetry breaking. We put down a set of rules to assign a spinor constraint with each non-zero charge. In all the configurations studied in this paper, they are shown to be the only spinor constraints contained in the KSEs.

In Section IV, we solve the KSEs of the toroidally compactified IIA superstring from Section III. Only the NS-NS charges are turned on in Section IVA. There are two sets of consistent KSEs, each one relates the spinors with opposite chiralities from 10-dimensional view point. The KSEs have identical structures with the toroidally compactified KSEs of heterotic string [22]. In Section IVB, we solve the KSEs with RR charges only. A solution with charges U-dual to the NS-NS charges of the configuration found in Secton IVA is explicitly obtained. It corresponds to the intersecting D-brane configuration with two D-2-branes and two D-4-branes in 10 dimensions. We show that the two configurations T-dual [37] to this solution are also solutions of the KSEs. The 10-dimensional interpretation of one of them is a configuration in which one D-0-brane is coupled to the intersection of three intersecting D-4-branes. The other configuration is a D-6-brane containing three intersecting D-2-branes. The classical configurations, composed of a large number of D-branes, obtained in this Section provides a consistency check of the D-brane intersection rules [36], which are defined microscopically, ${\it i.e.}$, in terms of a few D-branes only. In Section IVC, we find solutions with both NS-NS charges and RR charges. The first solution that we explicitly obtain corresponds to a bound state of a D-2-brane, a D-4-brane, and a fundamental string which lies on the intersection of the D-branes and carries a momentum, in a background with a magnetic monopole. We also show that the two configurations T-dual to the above configuration are also solutions of the KSEs. One of them corresponds to a bound state of a D-0-brane, a D-4-brane, a fundamental string which lies orthogonally to the D-4-brane and has a non-zero winding number, and a magnetic monopole, the gauge field of which associates with a toroidal direction orthogonal to both the D-4-brane and the fundamental string [47]. The other corresponds to a bound state of a D-6-brane, a D-2-brane, a solitonic 5-brane, and a fundamental string which lies on the intersection of the D-2-brane and the solitonic 5-brane and carries a momentum [40][48]. In all cases, the patterns of supersymmetry breaking are studied in detail. The BPS-saturated states with three to four charges preserve $N=1$ supersymmetry, those with two charges preserve $N=2$ supersymmetry, and those with one charge only preserve $N=4$ supersymmetry. Spinor constraints allow no more than four non-zero charges for the BPS-saturated states, under our assumptions of time-independence, spherical symmetry, and with dilaton and diagonal internal metric elements as the only non-zero scalars fields.

We make our conclusion in Section V.

\section{Effective action from 11-d supergravity on $T^7$ in IIA language}

In this section, we derive the field redefinition rules between the 4-d actions obtained from compactifying the N=1, d=11 SG on $T^7$ and that from N=2A, d=10 on $T^6$. That can simplify the way to obtain KSEs of the compactified IIA superstring in Section III. Most material in this section has been described in [26]. This section is included here for the sake of completeness and for establishing notations.

The field content of the $N$=1, d=11 SG consists of the following: Elfbein $E^{(11)\, A}_M$, gravitino $\psi^{(11)}_M$, and the 3-form field $A^{(11)}_{MNP}$.  The bosonic Lagrangian density is [28]

\begin{equation}
{\cal L} = -{1\over 4}E^{(11)}[{\cal R}^{(11)} + {1\over {12}}
F^{(11)}_{MNPQ}F^{(11)\ MNPQ} - {8\over {12^4}}
\varepsilon^{M_1 \cdots M_{11}}F_{M_1 \cdots M_4}F_{M_5 \cdots M_8}
A_{M_9 M_{10} M_{11}}] ,
\label{L11d}
\end{equation}
where $E^{(11)} \equiv {\rm det}\, E^{(11)\ A}_M$, ${\cal R}^{(11)}$ 
is the 11-dimensional Ricci scalar, and $F^{(11)}_{MNPQ} (\equiv 4\partial_{[M}A^{(11)}_{NPQ]})$ is the field strength associated with the 3-form field $A^{(11)}_{MNP}$. The metric signature is ($+--\cdots -$), and ($A,B,...$), ($M,N,...$) denote flat and curved indices in 11-d respectively.

Dimensional reduction of the 11-d SG to 4-d on $T^7$ is achieved by  
the following KK Ansatz for the Elfbein and a consistent truncation of the 
other 11-d fields:

\begin{equation}
E^{(11)\,A}_M = \left ( \matrix{e^{-{\varphi \over 2}} e^{\alpha}_{\mu} & 
B^i_{\mu} e^a_i \cr 0 & e^a_i} \right ) ,
\label{F11TO4}
\end{equation}
where $\varphi \equiv {\rm ln}\,{\rm det}\,e^a_i$, and $B^i_{\mu}$ 
($i=1,...,7$) are KK Abelian gauge fields.  Greek letters ($\alpha, \beta ,\cdots$) [($\mu ,\nu ,\cdots$)] are for the 4-d space-time flat [curved] indices while latin letters ($a,b,\cdots$) [($i,j,\cdots$)] are for the internal flat [curved] space indices. The 3-form field $A^{(11)}_{MNP}$ is truncated into three different types of 4-d fields: 35 pseudo-scalars $A_{ijk}$, 21 pseudo-vectors $A_{\mu\,ij}$ and 7 two-forms $A_{\mu\nu\,i}$. The two-forms $A_{\mu\nu\,i}$ are equivalent to (axionic) scalar fields $\varphi^i$ after making duality transformation.  In order to ensure that the fields $A_{\mu\,ij}$ and $A_{\mu\nu\,i}$ are scalars under the internal coordinate transformation $x^i \to x^{\prime\ i} = x^i + \xi^i$, and transform as $U(1)$ gauge fields under the gauge transformation: $\delta A^{(11)}_{MNP} = \partial_M \zeta_{NP} + \partial_N \zeta_{PM} + \partial_P \zeta_{MN}$, we have to define new canonical 4-d fields:

\begin{equation}
A^{\prime}_{\mu\,ij} \equiv A_{\mu\,ij} - B^k_{\mu} A_{kij},\ \ \ \
A^{\prime}_{\mu\nu\,i} \equiv A_{\mu\nu\,i} - B^j_{\mu}A_{j\nu\,i} 
-B^j_{\nu}A_{\mu\,ji} + B^j_{\mu} B^k_{\nu} A_{jki} .
\label{cantensor}
\end{equation}
The bosonic action (\ref{L11d}) is then reduced to the following effective 4-d action:

\begin{equation}
{\cal L} = -{1\over 4}e[{\cal R} - {1\over 2}\partial_{\mu} \varphi 
\partial^{\mu} \varphi +{1\over 4}\partial_{\mu} g_{ij} \partial^{\mu}
g^{ij} - {1\over 4} e^{\varphi}g_{ij}G^i_{\mu\nu}G^{j\,\mu\nu} 
+{1\over 2}e^{\varphi}g^{ik}g^{jl}F^4_{\mu\nu\,ij}
F^{4\,\mu\nu}_{\ \ \ \ kl} + \cdots ] ,
\label{L4df11d}
\end{equation}
where $e \equiv {\rm det}\,e^\alpha_\mu$, and ${\cal R}$ is the 4-d Ricci scalar. The Einstein-frame 4-d metric $g_{\mu\nu} = \eta_{\alpha\beta}e^{\alpha}_{\mu}e^{\beta}_{\nu}$, and $G^i_{\mu\nu} \equiv \partial_{\mu} B^i_{\nu} - \partial_{\nu}B^i_{\mu}$, $F^4_{\mu\nu\,ij} \equiv F^{\prime}_{\mu\nu\,ij} + G^k_{\mu\nu}A_{ijk}$. The dots ($\cdots$) denotes the terms involving the pseudo-scalars $A_{ijk}$ and the two-form fields $A_{\mu\nu\,i}$.  Here, $g_{ij} \equiv \eta_{ab}e^a_i e^b_j = -e^a_i e^a_j$ is the internal metric and the curved space indices ($i,j,...$) are raised by $g^{ij}$.

The zero slope limit of IIA 10-d superstring theory can be obtained by dimensional reduction of 11-d SG on a circle $S^1$ [27], with the following triangular gauge form for the Elfbein $E^{(11)\, A}_M$:

\begin{equation}
E^{(11)\, A}_M = \left ( \matrix{e^{-{\Phi \over 3}}
e^{(10)\, \breve{\alpha}}_{\breve{\mu}} & e^{{2\over 3}\Phi}
B_{\breve{\mu}} \cr 0 & e^{{2\over 3}\Phi}} \right ) ,
\label{F11TO10}
\end{equation}
where $\Phi$ corresponds to the 10-d dilaton field in Neveu-Schwarz-Neveu-Schwarz (NS-NS) sector of the superstring theory, $e^{(10)\, \breve{\alpha}}_{\breve{\mu}}$ is the Zehnbein in NS-NS sector, and $B_{\breve{\mu}}$ corresponds to a one-form in RR sector
\footnote{Therefore we are not assuming a diagonal 11-dimensional metric of the SG theory.}.  
Here, the breve denotes the 10-d space-time vector index. And the 3-form $A^{(11)}_{MNP}$ is decomposed into $A_{\breve{\mu}\breve{\nu}\breve{\rho}}$ and $A_{\breve{\mu}\breve{\nu} 11} (\equiv A_{\breve{\mu}\breve{\nu}})$, with $A_{\breve{\mu}\breve{\nu} \breve{\rho}}$ being identified as a Ramond-Ramond (RR) 3-form and $A_{\breve{\mu}\breve{\nu}}$ the NS-NS 2-form. The 11-d bosonic action (\ref{L11d}) becomes the following 10-d, $N=2$ SG action:

\begin{equation}
{\cal L} = {\cal L}_{NS} + {\cal L}_R , 
\label{N210d}
\end{equation}
with 

\begin{eqnarray}
{\cal L}_{NS} &=& -{1\over 4}e^{(10)}e^{-2\Phi}[{\cal R} + 
4\partial_{\breve{\mu}}\Phi \partial^{\breve{\mu}}\Phi - 
{1\over 3}F_{\breve{\mu}\breve{\nu}\breve{\rho}}
F^{\breve{\mu}\breve{\nu}\breve{\rho}}], 
\nonumber \\
{\cal L}_R &=& -{1 \over 4}e^{(10)}[{1\over 4}G_{\breve{\mu}\breve{\nu}}
G^{\breve{\mu}\breve{\nu}} +{1\over {12}}F^{\prime}_{\breve{\mu}\breve{\nu}
\breve{\rho}\breve{\sigma}} F^{\prime\, \breve{\mu}\breve{\nu}\breve{\rho}
\breve{\sigma}} - {6 \over {(12)^3}}\varepsilon^{\breve{\mu}_1 \cdots 
\breve{\mu}_{10}} F_{\breve{\mu}_1 \cdots \breve{\mu}_4}
F_{\breve{\mu}_5 \cdots \breve{\mu}_8} A_{\breve{\mu}_9 \breve{\mu}_{10}}],
\label{LNSR}
\end{eqnarray}
where $e^{(10)} \equiv {\rm det}\, e^{(10)\, \breve{\alpha}}_{\breve{\mu}}$, 
$\cal R$ is the 10-d string frame Ricci scalar, $F_{\breve{\mu}\breve{\nu}\breve{\rho}} \equiv 3\partial_{[\breve{\mu}}
A_{\breve{\nu}\breve{\rho}]}$, $G_{\breve{\mu}\breve{\nu}} \equiv 
2\partial_{[\breve{\mu}}B_{\breve{\nu}]}$, $F^{\prime}_{\breve{\mu}\breve{\nu}
\breve{\rho}\breve{\sigma}} \equiv 4\partial_{[\breve{\mu}}A_{\breve{\nu}
\breve{\rho}\breve{\sigma}]} - 4F_{[\breve{\mu}\breve{\nu}\breve{\rho}}
B_{\breve{\sigma}]}$, and $\varepsilon^{\breve{\mu}_1 \cdots \breve{\mu}_{10}} 
\equiv \varepsilon^{\breve{\mu}_1 \cdots \breve{\mu}_{10} 11}$.  
The ferminoic sector in 10-d contains Majorana gravitino $\psi_{\breve{\mu}}$ 
and fermion $\psi_{11}$ that come from the 11-d gravitino $\psi^{(11)}_M$, 
{\it i.e.}, $\psi^{(11)}_M = (\psi_{\breve{\mu}}, \psi_{11})$.  Each of these spinors can be splitted into two Majorana-Weyl spinors of two different chiralities.

In order to obtain the effective 4-d action of the IIA superstring 
compactified on $T^6$, we use the following KK Ansatz for the 
Zehnbein:
\begin{equation}
e^{(10)\, \breve{\alpha}}_{\breve{\mu}} = 
\left ( \matrix{e^{\alpha}_{\mu} & \bar{B}^m_{\mu}\bar{e}^a_m \cr 
0 & \bar{e}^a_m} \right ) , 
\label{F10TO4}
\end{equation}
where $\bar{B}^m_{\mu}$ ($m=1,...,6$) are Abelian KK gauge fields, 
$e^{\alpha}_{\mu}$ is the string frame 4-d Vierbein and $\bar{e}^a_m$ 
is the Sechsbein.   In the following, we set all the other scalars, 
except those associated with the Sechsbein $\bar{e}^a_m$ and the 10-d 
dilaton $\Phi$, to zero
\footnote{We turn off the scalar fields $B_{m}$ ($m$=4,...,9) associated 
with the 10-d $U(1)$ gauge field $B_{\breve{\mu}}$.  These fields are 
related to the internal metric coefficients $g_{m7}$ ($m=1,\cdots ,6)$ 
of 11-d SG.}.  

In this case, the string-frame 4-d bosonic action for IIA superstring is:

\begin{eqnarray}
{\cal L}_{II} = &-&{1\over 4}e[e^{-2\phi}({\cal R} + 4\partial_{\mu}
\phi \partial^{\mu}\phi + {1\over 4}\partial_{\mu}\bar{g}_{mn} 
\partial^{\mu}\bar{g}^{mn} - {1\over 4}\bar{g}_{mn}\bar{G}^m_{\mu\nu}
\bar{G}^{n\, \mu\nu} -\bar{g}^{mn}{\bar F}_{\mu\nu\, m}
{\bar F}^{\mu\nu}_{\ \ n}) 
\nonumber \\ 
&+&{1\over 4}e^{{\sigma}}\bar{G}_{\mu\nu}\bar{G}^{\mu\nu} + 
{1\over 2}e^{{\sigma}} \bar{g}^{mn}\bar{g}^{pq}
\bar{F}_{\mu\nu\, mp}\bar{F}^{\mu\nu}_{\ \ nq}] , 
\label{LIIAS}
\end{eqnarray}
where $e \equiv {\rm det}\, e^{\alpha}_{\mu}$, $2\phi \equiv 2\Phi - 
{\rm ln}\, {\rm det}\, \bar{e}^a_m$ (parameterizing the {\it string 
coupling}), ${\sigma} \equiv {\rm ln}\, {\rm det}\, \bar{e}^a_m$ 
(parameterizing the volume of 6-torus), $\bar{g}_{mn} \equiv \eta_{ab} 
\bar{e}^a_m \bar{e}^b_n = -\bar{e}^a_m \bar{e}^a_n$, and $\bar{G}^m_{\mu\nu} 
\equiv \partial_{\mu} \bar{B}^m_{\nu} - \partial_{\nu} \bar{B}^m_{\mu}$.
Here, the field strengths $\bar{F}_{\mu\nu\, m}$, $\bar{G}_{\mu\nu}$ and 
$\bar{F}_{\mu\nu\, mn}$ are defined in terms of the Abelian gauge fields 
decomposed from 10-d two-form $A_{\breve{\mu}\breve{\nu}}$, one-form 
$B_{\breve{\mu}}$ and the three-form $A_{\breve{\mu}\breve{\nu}
\breve{\rho}}$ fields, respectively. With the Weyl rescaling $g_{\mu\nu} \to g^E_{\mu\nu} = e^{-2\phi}g_{\mu\nu}$, we obtain the Einstein-frame action:

\begin{eqnarray}
{\cal L}_{II} = &-&{1\over 4}e^E[{\cal R}^E - 2\partial_{\mu}\phi 
\partial^{\mu} \phi + {1\over 4}\partial_{\mu}\bar{g}_{mn} 
\partial^{\mu}\bar{g}^{mn} - {1\over 4}e^{-2\phi}\bar{g}_{mn}
\bar{G}^m_{\mu\nu}\bar{G}^{n\,\mu\nu}-e^{-2\phi}\bar{g}^{mn}
{\bar F}_{\mu\nu\, m}{\bar F}^{\mu\nu}_{\ \ n} 
\nonumber \\ 
&+& {1\over 4}e^{\sigma}\bar{G}_{\mu\nu}\bar{G}^{\mu\nu} 
+ {1\over 2}e^{\sigma}\bar{g}^{mn}\bar{g}^{pq}
\bar{F}_{\mu\nu\, mp}\bar{F}^{\mu\nu}_{\ \ nq}],
\label{LIIAE}
\end{eqnarray}
where $e^E \equiv \sqrt{-{\rm det}\,g^E_{\mu\nu}}$ and ${\cal R}^E$ 
is the 4-d Einstein-frame Ricci scalar defined in terms of the metric $g^E_{\mu\nu}$.
 
As we have turned off the scalar fields associated with the 10-d 
$U(1)$ gauge field $B_{\breve{\mu}}$, ${\it i.e.}$, the internal metric
coefficients $g_{m7}$ of 11-d SG, the $SO(7)$ symmetry among the seven KK 
gauge fields and among the 21 3-form  gauge fields, separately, breaks down 
to the $SO(6)$ symmetry, which {\it do not mix} the gauge fields of RR 
and NS-NS sectors.  The RR sector consists of one KK gauge field 
$\bar{B}_\mu$, which transforms as a singlet of $SO(6)$, and fifteen 
3-form $U(1)$ gauge fields $\bar{A}_{\mu\, mn}$, which transform as 
{\bf 15} antisymmetric representation of $SO(6)$. The NS-NS sector 
consists of six KK gauge fields $\bar{B}^m_\mu$ and six 3-form $U(1)$ 
gauge fields $\bar{A}_{\mu\, n}$, each of them transform as a 
${\bf 6}$ vector representation of $SO(6)$
\footnote{In order to have the full manifestation of the $SO(7)$ symmetry of 11-d SG in the BH solutions of IIA superstring, the scalar fields which are associated with the 10-d $U(1)$ gauge field $B_{\breve{\mu}}$ has to be included.}. 

By keeping track of the field decomposition and redefinitions, and comparing the compactification Ans\" atze in the two schemes, {\it i.e.}, one corresponding to 11-d $\to$ 10-d $\to$ 4-d and the other one corresponding to 11-d $\to$ 4-d, wecan express the fields in the 4-d action of 11-d SG in terms of those in IIA superstring,

\begin{eqnarray}
\varphi &=& -{4 \over 3} \phi + {1 \over 3} \sigma,\ \ \   
{\rm ln} \, e_7^{\hat 7} = {2\over 3}\phi + {1\over 3} \sigma,\ \ \  
e_m^{\hat m} = e^{ -{1 \over 3} \phi - {1 \over 6} \sigma } {\bar{e}}_m^{\hat m}, 
\nonumber \\
B^m_{\mu} &=& {\bar{B}}^m_{\mu}, \ \ \ \  B_{\mu}^7 = \bar{B}_{\mu} , \ \ \ 
A_{\mu m n} = \bar{A}_{\mu m n}, \ \ \ \  A_{\mu\,m7} = \bar{A}_{\mu\,m}, 
\label{Rel}
\end{eqnarray}

where $m,n = 1,...,6$, flat indices are hatted, and the bar on Sechsbein is dropped.

\section{Killing Spinor Equations}

The supersymmetry transformation of the gravitino field $\psi^{(11)}_M$ 
in the $N=1$ 11-d theory (before compactification) in bosonic background is [28] 

\begin{equation}
\delta \psi^{(11)}_M = D_M\, \varepsilon +{i\over 144} (\Gamma^{NPQR}_
{\ \ \ \ \ M} - 8\Gamma^{PQR}\delta^N_M)F_{NPQR}\,\varepsilon ,
\label{STG11d}
\end{equation}
where $D_M\,\varepsilon = (\partial_M + {1\over 4}\Omega_{MAB}
\Gamma^{AB})\,\varepsilon$ is the gravitational covariant derivative 
on the spinor $\varepsilon$, and $\Omega_{ABC} \equiv -\tilde{\Omega}_
{AB,C} + \tilde{\Omega}_{BC,A} - \tilde{\Omega}_{CA,B}$ 
($\tilde{\Omega}_{AB,C} \equiv E^{(11)\,M}_{[A}E^{(11)\,N}_{B]}
\partial_N E^{(11)}_{MC}$) is the spin connection defined in terms of 
the Elfbein. With the Elfbein in (\ref{F11TO4}), the gravitino transformation (\ref{STG11d}) expressed in terms of 4-d canonical fields obtained by compactifying SG directly on $T^7$ are

\begin{eqnarray}
\delta \hat {\psi}_{\mu} &=& \partial_{\mu} \varepsilon + {1\over 4} 
\omega_{\mu\beta\gamma} \gamma^{\beta\gamma} \varepsilon - 
{1 \over 4}e^{\alpha}_{\mu}\eta_{\alpha[\beta}e^{\nu}_{\gamma ]}
\partial_{\nu} \varphi \gamma^{\beta\gamma} \varepsilon + 
{1 \over 8}(e^l_b \partial_{\mu}e_{lc} - e^l_c 
\partial_\mu e_{lb})\gamma^{bc} \varepsilon 
\nonumber \\
& &+ {i \over {24}}e^{\varphi \over 2} 
F_{\nu\rho\,ij}\gamma^{\nu\rho}_{\ \ \mu}\gamma^{ij} \varepsilon -
{i \over 6}e^{\varphi \over 2}F_{\mu\nu\,ij}
\gamma^{\nu}\gamma^{ij}\varepsilon + {1 \over 4} e^{\varphi \over 2} e_{ib} G^i_{\mu \alpha} \gamma^{\alpha 5} \gamma^b \varepsilon, 
\nonumber \\
\delta \psi_k &=& -{1\over 4}e^{\varphi \over 2}(\partial_{\rho} e_{kb} + 
e^c_k e^l_b \partial_{\rho} e_{lc})\gamma^{\rho 5} 
\gamma^b \varepsilon +{i \over {24}}e^{\varphi}F_{\mu\nu\,ij}
\gamma^{\mu\nu 5}\gamma^{ij}_{\ \  k}\varepsilon - {i\over 6}
e^{\varphi}F_{\mu\nu\,kl}\gamma^{\mu\nu 5}\gamma^l\varepsilon 
\nonumber \\
& &+ {1 \over 8} e^{\varepsilon} g_{kn} G^n_{\beta \alpha} \gamma^{\alpha \beta} \varepsilon
\label{STG411}
\end{eqnarray}
where $\delta \hat {\psi}_{\mu} \equiv \delta \psi_{\mu} - B^m_{\mu} \delta \psi_m$,  $\omega_{\mu\beta\gamma}$ is the spin-connection defined in terms 
of the Vierbein $e^{\alpha}_{\mu}$ and $[a\,\cdots\,b]$ denotes antisymmetrization of the corresponding indices.  For the 11-d gamma matrices, which satisfy the $SO(1,10)$ Clifford algebra $\{ \Gamma^A, \Gamma^B \} = 2\eta^{AB}$, we have used the following representation: 

\begin{equation}
\Gamma^{\alpha} = \gamma^{\alpha} \otimes I , \ \ \ \ \ \ 
\Gamma^a = \gamma^5 \otimes \gamma^a , 
\label{DFGAMMA}
\end{equation}
where $\{ \gamma^{\alpha}, \gamma^{\beta} \} = 2\eta^{\alpha\beta}$, 
$\{ \gamma^a , \gamma^b \} = -2\delta^{ab}$, $I$ is the $8 \times 8$ 
identity matrix and $\gamma^5 \equiv i\gamma^0 \gamma^1 \gamma^2 
\gamma^3$.  The above representation (\ref{DFGAMMA}) is compatible 
with the triangular gauge form (\ref{F11TO4}) ($SO(1,10) \to SO(1,3) 
\times SO(7)$) of the Elfbein. Multiple indices of the gamma matrices are antisymmetrized, {\it e.g.}, $\gamma^{\alpha\beta} \equiv \gamma^{[\alpha}\gamma^{\beta]}$, and the gamma matrices with curved indices are defined by multiplying with the Vierbein, {\it e.g.}, $\gamma^{\mu} \equiv e^{\mu}_{\alpha} \gamma^{\alpha}$. With the representation of the gamma matrices (\ref{DFGAMMA}), the spinor index $A$ of an 11-d spinor, $\varepsilon^A$, can be decomposed into $A = ({\bf a}, {\bf m})$, {\it i.e.}, $\varepsilon^A = \varepsilon^{({\bf a}, {\bf m})}$, where ${\bf a}=1,...,4$ is the spinor index for a four component 4-d spinor and ${\bf m}=1,...,8$ is the index for the spinor representation of the group $SO(7)$.

The supersymmetry transformations of the gravitinos and modulinos given in (\ref{STG411}) is a simple sum of the corresponding transformations from pure Kaluza-Klein sector, and pure 3-form fields sector [26]. That is not surprising, as the effective Lagrangian (\ref{L4df11d}) (after setting $A_{ijk}$ and $A_{\mu \nu\, i}$ to zero) is just a simple sum of the Ricci scalar, the kinetic energy of the scalar fields, and the kinetic energies of the two types of gauge fields, and contains no terms that describe any mixing between the two types of gauge fields. However, careful examination of the definitions of field strengths and tedious manipulation are required to verify explicitly the expected supersymmetry transformation rules.

We evaluate the KSEs in the following subsections under the assumptions of time-independence and spherical symmetry, and with the dilaton and diagonal internal metric elements as the only non-zero scalar fields. In subsection IIIA, the KSEs are expressed in terms of fields obtained by reducing SG directly on $T^7$. In subsection IIIB, the KSEs are expressed in terms of the fields from compactified IIA superstring. Then in subsection IIIC, we study the spinor constraints.

\subsection{From SG perspective}

In this subsection, we express the Killing spinor equations in terms of 4-dimensional fields from the compactified $N=1$ 11-dimensional SG. With spherical symmetry, the 4-d space-time metric can be taken as 

\begin{equation}
g_{\mu\nu}{\rm d}x^{\mu}{\rm d}x^{\nu} = \lambda (r){\rm d}t^2
- \lambda^{-1} (r) {\rm d}r^2 - R(r)({\rm d}\theta^2 + {\rm sin}^2 \theta
{\rm d}\phi^2) .
\label{MET4d}
\end{equation}
Field strengths $G_{\mu \nu}^i$ and $F_{\mu \nu ij}$ for Kaluza-Klein $U(1)$ gauge field and that from three-form fields, respectively, have the following non-zero components:

\begin{eqnarray}
G^i_{tr} = {{g^{ij} Q_j} \over {R e^{\varphi}}}, \ \ \ \ \ 
G^i_{\theta \phi} = P^i {\rm sin} \theta ,
\nonumber \\
F_{trij} = {{g_{ik} g_{jl} Q^{kl}} \over {R e^{\varphi}}}, \ \ \ \ \ 
F_{\theta \phi ij} = P_{ij} {\rm sin} \theta ,
\label{PQ}
\end{eqnarray}
where $Q_i$ ($P^i$) and $Q^{ij}$ ($P_{ij}$) are the physical electric (magnetic) charges
\footnote{To simplify our formulae, we have assumed that the internal metric and the dilaton approach unity and zero respectively as r $\rightarrow \infty$. This can always be done through manifest $SL(7,R)$ symmetry and S-duality.}.
The internal metric, $g_{ij}$, is proportional to $\delta_{ij}$ by assumption. The supersymmetry transformations of the gravitinos and modulinos, (\ref{STG411}), can thus be simplified and written as:

\begin{equation}
{i \over 4} \left( - {\bf P}^m \mp i {\bf Q}_m \right) \varepsilon_{ul} - {1 \over 2} R \sqrt{\lambda} ( \ln e_m^{\hat m} )' \gamma^m \varepsilon_{lu} + {1 \over 12} ( \pm {\bf P}_{ij} - i {\bf Q}^{ij} ) \gamma^{ij}_{\ \ m} \varepsilon_{ul} + {1 \over 3} ( \mp {\bf P}_{mi} + i {\bf Q}^{mi} ) \gamma^i \varepsilon_{ul} = 0 ,
\label{DSM11}
\end{equation}

\begin{equation}
\mp {R \over \sqrt{\lambda}} \left( \lambda ' - \lambda \varphi ' \right) \varepsilon_{ul} - {\bf Q}_i \gamma^i \varepsilon_{lu} + {1 \over 3}( {\bf P}_{ij} \mp 2 i {\bf Q}^{ij} ) \gamma^{ij} \varepsilon_{lu} = 0 ,
\label{DST11}
\end{equation}

\begin{equation}
{i \over 2} \sqrt{R} \varepsilon_{ul} - {i \over 4} \sqrt{\lambda} ( R' - R \varphi ' ) \varepsilon_{ul} + {1 \over 4} {\bf P}^i \gamma^i \varepsilon_{lu} + {1 \over 12} ( - {\bf Q}^{ij} \mp 2 i {\bf P}_{ij} ) \gamma^{ij} \varepsilon_{lu} = 0 ,
\label{DSTH11}
\end{equation}

\begin{equation}
R \sqrt{\lambda} \partial_r \varepsilon_{ul} \pm {1 \over 4} {\bf Q}_i \gamma^i \varepsilon_{lu} + {1 \over 12}( \mp {\bf P}_{ij} + 2 i {\bf Q}^{ij} ) \gamma^{ij} \varepsilon_{lu} = 0 ,
\label{DSR11}
\end{equation}

\begin{equation}
(\varepsilon^{1, {\bf m}}_{u, \ell}, \varepsilon^{2, {\bf m}}_{u,\ell}) = 
e^{i\sigma^2 \theta /2} e^{i\sigma^3 \phi /2}(a^{1, {\bf m}}_{u,\ell}(r), 
a^{2, {\bf m}}_{u,\ell}(r)) , 
\label{DSTHS11}
\end{equation}
where

\begin{eqnarray}
{\bf{Q}}_m \equiv e^{- {\varphi \over 2}} e^m_{\hat m} Q_m,\ \ \ \ \ {\bf{P}}^m \equiv e^{\varphi \over 2} e_m^{\hat m} P_m ,
\nonumber \\
{\bf{Q}}^{mi} \equiv e^{- {\varphi \over 2}} e_m^{\hat m} e_i^{\hat i} Q^{mi}, \ \ \ \ \ {\bf{P}}_{mi} \equiv e^{\varphi \over 2} e^m_{\hat m} e^i_{\hat i} P_{mi} ,
\label{DFBPQ11}
\end{eqnarray}
and $\varepsilon^{\bf m}_{u, \ell}$ are the upper (or lower) two components of the 4-d four-component spinor, $\varepsilon^{\bf m}$, {\it i.e.}, $(\varepsilon^{\bf m})^T = (\varepsilon^{\bf m}_u, \varepsilon^{\bf m}_\ell )$, and $a^{\bf m}_{u,\ell}(r)$ are the corresponding two-component spinors
\footnote{We also call the quantities, $\varepsilon^{\bf m}_u$ and $\varepsilon^{\bf m}_l$, as two-component ${\it spinors}$. That only mean that they are the upper and lower two components of the 4-d spinors, $\varepsilon^{\bf m}$, respectively. Actually, the 4-d spinors have to be Majorana (shown in the Appendix), therefore the upper and lower components do not have definite transformation properties under the 4-d Lorentz group.}
that depend on the radial coordinate $r$ only
\footnote{Note that we suppress the index ${\bf m}$ of the spinors in all the equations in this paper.}.
The KSEs (\ref{DSR11}) and (\ref{DSTHS11}) determine the radial and angular dependence of the spinors. As all information of the fields and the constraints on the spinors are contained in (\ref{DSM11}) to (\ref{DSTH11}), we shall not elaborate equations (\ref{DSR11}) and (\ref{DSTHS11}) any further (most of the above technical details were developed in [29] and [26]).

Note that the gamma matrices in the KSEs only act on the index $\bf m$, which is the index for the spinor representation of the group $SO(7)$ used in reducing the spinor in 11-d SG on $T^7$, (\ref{DFGAMMA}). We have used explicit representation of the 4-d gamma matrices, corresponding to the spinor representation of the 4-d space-time Lorentz group, to write the KSEs in terms of relations between upper and lower components of the 4-d spinors. 

\subsection{From IIA perspective}

In order to rewrite the above KSEs, (\ref{DSM11})-(\ref{DSTH11}), in IIA language, we define the projection operators

\begin{equation}
P_+ \equiv {1 \over 2} \left( 1 - i \gamma^7 \right), \ \ \ \ \
P_- \equiv {1 \over 2} \left( 1 + i \gamma^7 \right),
\label{PP}
\end{equation}
and get the two projections from $\varepsilon$,

\begin{equation}
\varepsilon = \varepsilon^+ + \varepsilon^- ,
\label{ESPM}
\end{equation}
where
\begin{equation}
\varepsilon^+ \equiv P_+ \varepsilon,\ \ \ \ \ \varepsilon^- \equiv P_- \varepsilon .
\label{DFESPM}
\end{equation}
The two types of spinors, $\varepsilon^{{\bf m}+}$ and $\varepsilon^{{\bf m}-}$, originate from the two spinors associated with the $N=2$ supersymmetry of the IIA superstring in 10 dimensions
\footnote{In the Appendix, we show that the chirality of a 10-d Majorana-Weyl spinor is labelled by the corresponding eigenvalue of $\gamma^7$ which only acts on the index ${\bf m}$, when the 10-d Lorentz group is considered as a direct product of the 4-d Lorentz group and the 6-d rotation group.}.

With (\ref{Rel}), the SG charges defined in (\ref{DFBPQ11}) can be written in terms of charges from NS-NS sector and RR sector of the IIA superstring as

\begin{eqnarray}
{\bf{Q}}_i = ( 1 - \delta^7_i ) {\bf{Q}}^{NK}_i - \delta^7_i {\bf{Q}}^{RK},\ \ \ \ \ {\bf{P}}^i = ( 1 - \delta^7_i ) {\bf{P}}^i_{NK} + \delta^7_i {\bf{P}}_{RK}
\nonumber \\
{\bf{Q}}^{ij} = ( 1 - \delta^7_i ) ( 1 - \delta^7_j ) {\bf{Q}}^{ij}_{RF} - ( 1 - \delta^7_i ) \delta^7_j {\bf{Q}}^i_{NF} + ( 1 - \delta^7_j ) \delta^7_i {\bf{Q}}^j_{NF}
\nonumber \\
{\bf{P}}_{ij} = ( 1 - \delta^7_i ) ( 1 - \delta^7_j ) {\bf{P}}_{ij}^{RF} + ( 1 - \delta^7_i ) \delta^7_j {\bf{P}}_i^{NF} - ( 1 - \delta^7_j ) \delta^7_i {\bf{P}}_j^{NF}
\label{PQ11NSR}
\end{eqnarray}
where

\begin{eqnarray}
{\bf{Q}}^{NK}_i \equiv e^{\phi} \bar{e}^i_{\hat i} Q^{NK}_i, \ \ \ \ \ {\bf{Q}}^{RK} \equiv e^{- {1 \over 2} \sigma} Q^{RK}
\nonumber \\
{\bf{P}}_{NK}^i \equiv e^{- \phi} \bar{e}_i^{\hat i} P_{NK}^i, \ \ \ \ \ {\bf{P}}_{RK} \equiv e^{{1 \over 2} \sigma} P_{RK}
\nonumber \\
{\bf{Q}}^{i}_{NF} \equiv e^{\phi} \bar{e}_i^{\hat i} Q_{NF}^i, \ \ \ \ \ {\bf{Q}}^{ij}_{RF} \equiv e^{- {1 \over 2} \sigma} \bar{e}^{\hat i}_i \bar{e}^{\hat j}_j Q^{ij}_{RF}
\nonumber \\
{\bf{P}}_{i}^{NF} \equiv e^{- \phi} \bar{e}^i_{\hat i} P^{NF}_i, \ \ \ \ \ {\bf{P}}_{ij}^{RF} \equiv e^{ {1 \over 2} \sigma} \bar{e}_{\hat i}^i \bar{e}_{\hat j}^j P^{RF}_{ij}
\label{PQNSR}
\end{eqnarray}
and $N, R, K, F$ indicate that the charge is from NS-NS sector, RR sector, with Kaluza-Klein origin, and with 1-form, 3-form (in RR sector) or 2-form (in NS-NS sector) origin respectively. The index, $i$, indicates that the charge is associated with the $i$ th compactified dimension. From (\ref{PQ}), $G^7_{tr} \rightarrow - {{Q_7} \over r^2}$ as $g_{77} \rightarrow -1$ asymtotically. As ${\bar B}_{\mu} \equiv B^7_{\mu}$ from (\ref{Rel}), and $Q^{RK}$ is defined to be the charge of the gauge field ${\bar B}_{\mu}$, ${\it i.e.}$ ${\bar G}_{tr} \rightarrow {Q^{RK} \over r^2}$, therefore, $Q^{RK} = -Q_7$. Similarly, $Q^{NF}_i = - Q^{i7}$ as $F_{tri7}$ contains the extra factor, $g_{77}$, compared with the definition of $Q^i_{NF}$. No such flipping of signs are needed for the magnetic charges, as the internal metric is not involved in their definition. 

With equations (\ref{Rel}) and (\ref{PQ11NSR}), we act $P_+, P_-$ on both sides of (\ref{DSM11}) and get
\footnote{We shall normalize the charges such that $(Q_{NF}, P^{NF}, Q_{RF}, P^{RF}) \rightarrow {1 \over 2} (Q_{NF}, P^{NF}, Q_{RF}, P^{RF})$, to make the equations more symmetrical.}:

\begin{eqnarray}
- R \sqrt{\lambda} ( \ln \bar{e}^{\hat{m}}_m )' \varepsilon^+_{ul} =
{1 \over 2} \left[ \left( \mp {\bf{Q}}^{NK}_m - {\bf{Q}}^m_{NF} \right) + i \left( - {\bf{P}}^m_{NK} \mp {\bf{P}}^{NF}_m \right) \right] \gamma^m \varepsilon^-_{lu} 
\nonumber \\
+ {1 \over 4} \left[ {1 \over 2} \left( \mp {\bf{P}}^{RF}_{ij} - i {\bf{Q}}^{ij}_{RF} \right) \gamma^{ij} + 2 \left( \pm {\bf{P}}^{RF}_{mi} + i {\bf{Q}}^{mi}_{RF} \right) \gamma^{mi} + {\bf{P}}_{RK} \pm i {\bf{Q}}^{RK} \right] \varepsilon^+_{lu}
\nonumber 
\end{eqnarray}

\begin{eqnarray}
- R \sqrt{\lambda} ( \ln \bar{e}^{\hat{m}}_m )' \varepsilon^-_{ul} =
{1 \over 2} \left[ \left( \mp {\bf{Q}}^{NK}_m + {\bf{Q}}^m_{NF} \right) + i \left( - {\bf{P}}^m_{NK} \pm {\bf{P}}^{NF}_m \right) \right] \gamma^m \varepsilon^+_{lu} 
\nonumber \\
+ {1 \over 4} \left[ {1 \over 2} \left( \mp {\bf{P}}^{RF}_{ij} - i {\bf{Q}}^{ij}_{RF} \right) \gamma^{ij} + 2 \left( \pm {\bf{P}}^{RF}_{mi} + i {\bf{Q}}^{mi}_{RF} \right) \gamma^{mi} - {\bf{P}}_{RK} \mp i {\bf{Q}}^{RK} \right] \varepsilon^-_{lu} ,
\label{NSRM}
\end{eqnarray}
where $m =$ 1,...6. From $\delta \hat {\psi}_7 = 0$, 

\begin{eqnarray}
R \sqrt{\lambda} ( \phi + {1 \over 2} \sigma )' \varepsilon^+_{ul} =
{1 \over 4} \left[ 3 \left( - {\bf{P}}_{RK} \mp i {\bf{Q}}^{RK} \right) + {1 \over 2} \left( \pm {\bf{P}}^{RF}_{ij} + i {\bf{Q}}^{ij}_{RF} \right) \gamma^{ij} \right] \varepsilon^+_{lu} 
\nonumber \\
+ {1 \over 2} \left( \pm i {\bf{P}}^{NF}_i + {\bf{Q}}^{i}_{NF} \right) \gamma^{i} \varepsilon^-_{lu}
\nonumber 
\end{eqnarray}

\begin{eqnarray}
R \sqrt{\lambda} ( \phi + {1 \over 2} \sigma )' \varepsilon^-_{ul} =
{1 \over 4} \left[ -3 \left( - {\bf{P}}_{RK} \mp i {\bf{Q}}^{RK} \right) + {1 \over 2} \left( \pm {\bf{P}}^{RF}_{ij} + i {\bf{Q}}^{ij}_{RF} \right) \gamma^{ij} \right] \varepsilon^-_{lu}
\nonumber \\
 - {1 \over 2} \left( \pm i {\bf{P}}^{NF}_i + {\bf{Q}}^{i}_{NF} \right) \gamma^{i} \varepsilon^+_{lu} ,
\label{NSR7}
\end{eqnarray}
With (\ref{NSR7}), we can write (\ref{DST11}), (\ref{DSTH11}), respectively, as 

\begin{eqnarray}
R \sqrt{\lambda} \left( \ln {\lambda} + 2 {\phi} \right)' \varepsilon^+_{ul} = \left( \mp {\bf Q}^{NK}_i + {\bf Q}^i_{NF} \right) \gamma^i \varepsilon^-_{lu} + {1 \over 4} \left[ \left( \pm {\bf P}^{RF}_{ij} - i {\bf Q}^{ij}_{RF} \right) \gamma^{ij} + 2 \left( - {\bf P}_{RK} \pm i {\bf Q}^{RK} \right) \right] \varepsilon^+_{lu}
\nonumber \\
R \sqrt{\lambda} \left( \ln {\lambda} + 2 {\phi} \right)' \varepsilon^-_{ul} = \left( \mp {\bf Q}^{NK}_i - {\bf Q}^i_{NF} \right) \gamma^i \varepsilon^+_{lu} + {1 \over 4} \left[ \left( \pm {\bf P}^{RF}_{ij} - i {\bf Q}^{ij}_{RF} \right) \gamma^{ij} - 2 \left( - {\bf P}_{RK} \pm i {\bf Q}^{RK} \right) \right] \varepsilon^-_{lu} ,
\label{NSRT}
\end{eqnarray}

\begin{eqnarray}
R \sqrt{\lambda} \left( \ln {\lambda} - 2 {\phi} \right)' \varepsilon^+_{ul} =i  \left( {\bf P}_{NK}^i \mp {\bf P}_i^{NF} \right) \gamma^i \varepsilon^-_{lu} + {1 \over 4} \left[ \left( \pm {\bf P}^{RF}_{ij} - i {\bf Q}^{ij}_{RF} \right) \gamma^{ij} + 2 \left( - {\bf P}_{RK} \pm i {\bf Q}^{RK} \right) \right] \varepsilon^+_{lu}
\nonumber \\
R \sqrt{\lambda} \left( \ln {\lambda} - 2 {\phi} \right)' \varepsilon^-_{ul} =i  \left( {\bf P}_{NK}^i \pm {\bf P}_i^{NF} \right) \gamma^i \varepsilon^+_{lu} + {1 \over 4} \left[ \left( \pm {\bf P}^{RF}_{ij} - i {\bf Q}^{ij}_{RF} \right) \gamma^{ij} - 2 \left( - {\bf P}_{RK} \pm i {\bf Q}^{RK} \right) \right] \varepsilon^-_{lu} ,
\label{NSRTH}
\end{eqnarray}
Therefore NS-NS charges relate spinors with different chiralities (from 10-d view point), while RR charges relate spinors with the same chirality.

\subsection{Spinor constraints}

We would like to state the spinor constraints associated with the charges in the KSEs in this subsection. The patterns of supersymmetry breaking are governed by the spinor constraints. Each spinor constraint reduces half of the spinor degree of freedom, leading to the breaking of half of the supersymmetry. From the KSEs, (\ref{NSRM}) to (\ref{NSRTH}), we see that we can associate a spinor constraint to each charge such that the consistency of the KSEs is guaranteed when only that charge is non-zero. For example, if only $Q^{NK}_m$ is non-zero, there are two sets of consistent KSEs. One set relates $\varepsilon^+_u$ with $\varepsilon^-_l$, with spinor constraint: $\eta_* \varepsilon^+_u = \gamma^m \varepsilon^-_l$, where $\eta_*$ = $\pm 1$. The other set relates $\varepsilon^-_u$ with $\varepsilon^+_l$, with spinor constraint: $\eta_* \varepsilon^-_u = \gamma^m \varepsilon^+_l$. 

Here we collect all the eight spinor constraints associated with the eight different types of charges: 

\begin{eqnarray}
Q^{NK}_m: \ \ \ \ \ \eta_1 \varepsilon^+_u = \gamma^m \varepsilon^-_l, \ \ \ \ \ \eta_1 \varepsilon^-_u = \gamma^m \varepsilon^+_l,
\nonumber
\end{eqnarray}
\begin{eqnarray}
Q^{NF}_m: \ \ \ \ \ \eta_2 \varepsilon^+_u = \gamma^m \varepsilon^-_l, \ \ \ \ \ \eta_2 \varepsilon^-_u = - \gamma^m \varepsilon^+_l,
\nonumber
\end{eqnarray}
\begin{eqnarray}
P^{NK}_m: \ \ \ \ \ \eta_3 \varepsilon^+_u = i \gamma^m \varepsilon^-_l, \ \ \ \ \ \eta_3 \varepsilon^-_u = i \gamma^m \varepsilon^+_l,
\nonumber
\end{eqnarray}
\begin{eqnarray}
P^{NF}_m: \ \ \ \ \ \eta_4 \varepsilon^+_u = i \gamma^m \varepsilon^-_l, \ \ \ \ \ \eta_4 \varepsilon^-_u = -i \gamma^m \varepsilon^+_l,
\nonumber
\end{eqnarray}
\begin{eqnarray}
Q^{RF}_{ij}: \ \ \ \ \ \eta_5 \varepsilon^+_u = i \gamma^{ij} \varepsilon^+_l, \ \ \ \ \ \eta_5 \varepsilon^-_u = i \gamma^{ij} \varepsilon^-_l,
\nonumber
\end{eqnarray}
\begin{eqnarray}
P^{RF}_{ij}: \ \ \ \ \ \eta_6 \varepsilon^+_u = \gamma^{ij} \varepsilon^+_l, \ \ \ \ \ \eta_6 \varepsilon^-_u = \gamma^{ij} \varepsilon^-_l,
\nonumber
\end{eqnarray}
\begin{eqnarray}
Q^{RK}: \ \ \ \ \ \eta_7 \varepsilon^+_u = - i \varepsilon^+_l, \ \ \ \ \ \eta_7 \varepsilon^-_u = i \varepsilon^-_l,
\nonumber
\end{eqnarray}
\begin{equation}
P^{RK}: \ \ \ \ \ \eta_8 \varepsilon^+_u = - \varepsilon^+_l, \ \ \ \ \ \eta_8 \varepsilon^-_u = \varepsilon^-_l .
\label{GSC}
\end{equation}
where $\eta_i = \pm 1$. The bosonic configurations in general depend on the $\eta$s. 

In all of the configurations we considered in this paper (each has more than one charge), we find that explicit evaluation of the KSEs does give a spinor constraint associated with each non-zero charge, and the form of the constraint is precisely the same as in (\ref{GSC})
\footnote{One interesting implication of these spinor constraints is that there is no static, spherically symmetric BPS-saturated configuration that contains both $P^{RK}$ and $P^{NK}$, or both $P^{RK}$ and $Q^{NF}$, with all scalar fields vanish except the dilaton and the diagonal internal metric elements. }.

\section{BPS configurations}

In this section, we find explicit solutions to the KSEs. In Section IVA, only NS-NS charges are turned on, the RR charges are turned off. Then in Section IVB, only RR charges are turned on. In Section IVC, configurations with charges from both NS-NS and RR sectors are discussed.

\subsection{Neveu-Schwarz-Neveu-Schwarz sector}

With the RR charges set to zero, we find, from the KSEs (\ref{NSRM}) - (\ref{NSRTH}), two sets of consistent KSEs relating the spinors: ${\varepsilon}^+_l$ to ${\varepsilon}^-_u$, and ${\varepsilon}^+_u$ to ${\varepsilon}^-_l$. These two sets of KSEs can be written together as follows:

\begin{equation}
-R \sqrt{\lambda} ( \ln \bar{e}^{\hat{m}}_m )' \varepsilon_{ul} =
{1 \over 2} \left[ \pm \left( \eta_* {\bf{Q}}^{NK}_m - {\bf{Q}}_m^{NF} \right) + i \left( -{\bf{P}}_m^{NK} + \eta_* {\bf{P}}^{NF}_m \right) \right] \gamma^m \varepsilon_{lu}
\label{HETME}
\end{equation}

\begin{equation}
R \sqrt{\lambda} \left( {\phi} + {1 \over 2} \sigma \right)' \varepsilon_{ul} =
{1 \over 2} \left( - i \eta_* {\bf{P}}^{NF}_i \pm {\bf{Q}}_{i}^{NF} \right) \gamma^{i} \varepsilon_{lu}
\label{HETD}
\end{equation}

\begin{equation}
R \sqrt{\lambda} \left( \ln {\lambda} + 2 {\phi} \right)' \varepsilon_{ul} = \pm \left( \eta_* {\bf Q}^{NK}_i + {\bf Q}_i^{NF} \right) \gamma^i \varepsilon_{lu}
\label{HETP}
\end{equation}

\begin{equation}
R \sqrt{\lambda} \left( \ln {\lambda} - 2 {\phi} \right)' \varepsilon_{ul} =i  \left( {\bf P}^{NK}_i + \eta_* {\bf P}_i^{NF} \right) \gamma^i \varepsilon_{lu}, \label{HETM}
\end{equation}
where $\eta_*$ = 1 when $({\varepsilon}_u, {\varepsilon}_l) \equiv ({\varepsilon}^+_l, {\varepsilon}^-_u)$, and $\eta_*$ = -1 when $({\varepsilon}_u, {\varepsilon}_l) \equiv ({\varepsilon}^+_u, {\varepsilon}^-_l)$. Equation (\ref{HETD}) originates from (\ref{NSR7}) when all RR charges are turned off. Therefore the two components of $\delta \hat {\psi}_{7}$ with opposite chiralities from 10-d view point can be identified with the two dilatinos of the IIA superstring in 10 dimensions. 

The structure of this set of KSEs is identical to that of heterotic string [22]. Actually, we can reproduce the KSEs of the toroidally compactified heterotic string with the above KSEs of the toroidally compactified IIA superstring at each $\eta_*$ with the following maps:

\begin{equation}
\varepsilon^+_l \rightarrow \varepsilon_u, \ \ \ \
\varepsilon^-_u \rightarrow \varepsilon_l, \ \ \ \, \eta_* = 1,
\label{MAP1}
\end{equation}

\begin{equation}
\varepsilon^+_u \rightarrow \varepsilon_u, \ \ \ \
\varepsilon^-_l \rightarrow \varepsilon_l, \ \ \ \, \eta_* = -1,
\label{MAP2}
\end{equation}

\begin{equation}
\eta_* Q^{NK}_m \rightarrow Q^{(1)}_m, \ \ \ \ 
\eta_* P^{NF}_m \rightarrow P^{(2)}_m, \ \ \ \, \eta_* = \pm 1,
\label{MAP12}
\end{equation}
where the quantities on the left of $\rightarrow$ belong to the compactified IIA superstring, and the right quantities belong to the compactified heterotic string. The superscript, 1 and 2, of the charges from heterotic string indicates the origin, ${\it i.e.}$, Kaluza-Klein gauge fields or two-form fields, respectively.

In the case of heterotic string, which has $N=1$ in 10d and $N=4$ in 4d, the KSEs relate the upper and lower components of the 4-d spinors which originate from the same 10-d Majorana-Weyl spinor with a definite chirality. While in the case of IIA superstring, which has $N=2$ in 10d and $N=8$ in 4d, the KSEs expressed in terms of 4-d fields relate the upper(lower) components of the 4-d spinors which originate from a 10-d Majorana-Weyl spinor of certain chirality to the lower(upper) components of spinors which originate from another 10-d Majorana-Wely spinor of opposite chirality. 

From (\ref{HETP}), non-zero $( \eta_* {\bf Q}_i^{NK} + {\bf Q}_i^{NF})$ gives a spinor constraint of the form : $\varepsilon_l = - \eta_* \eta_+ \gamma^i \varepsilon_u$. From (\ref{HETM}), non-zero $( {\bf P}_j^{NK} + \eta_* {\bf P}_j^{NF})$ gives a spinor constraint of the form : $\varepsilon_l = -i \eta_- \gamma^j \varepsilon_u$, where $\eta_{\pm}$ can be equal to $+1$ or $-1$. Therefore the maximum number of charges allowed by constraints on spinors is four, with one electric NK and one electric NF charge from the same compactified dimension, and one magnetic NK and one magnetic NF charge from another compactified dimension
\footnote{We follow the same line of argument used in finding the generating solution for the supersymmetric, spherically symmetric solutions in Abelian Kaluza-Klein theory [29].}.
Without loss of generality, we choose the non-zero charges to be: $P^{NK}_1, P^{NF}_1, Q^{NK}_2, Q^{NF}_2$. Solving (\ref{HETME}) to (\ref{HETM}), we get the fields:

\begin{eqnarray}
{ \lambda = { r^2 \over { \left[ ( r + \eta_p P^{NK}_1 ) ( r + \eta_* \eta_p P^{NF}_{1} ) ( r + \eta_q Q^{NK}_{2} ) ( r + \eta_* \eta_q Q^{NF}_{2} ) \right]^{1 \over 2} }}}
\nonumber 
\end{eqnarray}

\begin{eqnarray}
{ R = \left[ ( r + \eta_p P^{NK}_1 ) ( r + \eta_* \eta_p P^{NF}_{1} ) ( r + \eta_q Q^{NK}_{2} ) ( r + \eta_* \eta_q Q^{NF}_{2} ) \right]^{1 \over 2} }
\nonumber 
\end{eqnarray}

\begin{eqnarray}
e^{2 \phi} = \left[ { { (r+\eta_p P^{NK}_1) (r+ \eta_* \eta_p P^{NF}_1) } \over { (r+\eta_q Q^{NK}_2) (r+ \eta_* \eta_q Q^{NF}_2) } } \right]^{1 \over 2}
\nonumber 
\end{eqnarray}

\begin{eqnarray}
e^{\sigma} = \left[ { { (r+\eta_q Q^{NK}_2) (r+ \eta_* \eta_p P^{NF}_1) } \over { (r+\eta_* \eta_q Q^{NF}_2) (r+\eta_p P^{NK}_1) } } \right]^{1 \over 2}
\nonumber 
\end{eqnarray}

\begin{eqnarray}
e^{\hat 1}_1 = \left( { {r+\eta_* \eta_p P^{NF}_1} \over {r+ \eta_p P^{NK}_1} } \right)^{1 \over 2}
\nonumber 
\end{eqnarray}

\begin{eqnarray}
e^{\hat 2}_2 = \left( { {r+ \eta_q Q^{NK}_2} \over {r+ \eta_* \eta_q Q^{NF}_2} } \right)^{1 \over 2}
\nonumber 
\end{eqnarray}

\begin{equation}
e^{\hat m}_m = 1, \ \ \ \ \ m = 3,...,6 ,
\label{HETSOL}
\end{equation}
The radial distance is defined such that the horizon is at $r=0$. The $\eta$s (i.e., $\eta_+, \eta_-, \eta_*$ ) can be equal to $\pm 1$. Note that we have dropped the bars on the internal metric components.

This configuration has a 10-dimensional interpretation as a fundamental string lying within the solitonic 5-brane [23], and bounded with a magnetic monople. The string wraps around the 2nd toroidal direction and has non-zero winding number, thereby giving the charge, $Q^{NF}_2$. It carries a momentum and so provides the charge, $Q^{NK}_2$. The 5-brane wraps around the toroidal directions, (23456), and carries the charge, $P^{NF}_1$. The magnetic monople carries the charge, $P^{NK}_1$.

For the spinor constraints, each non-zero magnetic charge ($P^{NK}_1$, $P^{NF}_1$, or both) and each non-zero electric charge ($Q^{NK}_2$, $Q^{NF}_2$, or both) respectively creates a constraint 

\begin{equation}
\varepsilon_u = i \eta_p \gamma^1 \varepsilon_l, \ \ \ \ \ \
\varepsilon_u = \eta_* \eta_q \gamma^2 \varepsilon_l .
\label{NSSC}
\end{equation}
Note that these constraints are expected from (\ref{GSC}) with suitable definitions of the $\eta$s. 

The mass of the black hole from (\ref{HETSOL}) is

\begin{equation}
M = {1 \over 4} \left[ \eta_- \left( P^{NK}_1 + \eta_* P^{NF}_1 \right) + \eta_+ \left( Q^{NK}_2 + \eta_* Q^{NF}_2 \right) \right]
\label{NSMASS}
\end{equation}
The eight different combinations of $\eta$s correspond to the positive and negative values of the four central charges [30][31] : $|Q_R + P_R|, |Q_R - P_R|, |Q_L + P_L|, |Q_L - P_L|$, where $P_{R,L} \equiv P^{NK}_1 \pm P^{NF}_1$ and $Q_{R,L} \equiv Q^{NK}_2 \pm Q^{NF}_2$. Mass of a BPS state is equal to the maximum of the central charges, thus $\eta$s have to be chosen to maximize the mass given by (\ref{NSMASS}). Consequently there is no massless solution unless all charges vanish, i.e., no gauge or supersymmetry enhancement. This is in contrast with the case of $N=4$ heterotic string [16]. The difference is made by the extra freedom of maximizing the mass with $\eta_*$, which is a result of the $N=8$ supersymmetry.

With three $\eta$s and four non-zero charges, there always exists the possibility of getting a solution with naked singularity. For example, a configuration with the charges that satisfy the inequalities: $P^{NK}_1, P^{NF}_1, Q^{NK}_2 \gg -Q^{NF}_2 > 0$, have all the $\eta$s equal to one in order to maximize the mass. Therefore there is a naked singularity at $r = - Q^{NF}_2 $ from (\ref{HETSOL}). For a regular solution (${\it e.g.}$, when all four charges are positive), the mass is equal to the sums of the four absolute values of the charges, ${\it i.e.}$, 

\begin{equation}
M = {1 \over 4} \left( | P^{NK}_1 | + | P^{NF}_1 | + | Q^{NK}_2 | + | Q^{NF}_2 | \right) .
\label{RNSMASS}
\end{equation}
When only three or fewer NS-NS charges are non-zero, the three $\eta$s are chosen in such a way that the mass is proportional to the sum of the absolute values of the charges. The horizon coincides with the singular surface of the black hole, and consequently the black hole has zero entropy. 

With the KSEs explicitly solved, we can study the pattern of supersymmetry breaking in detail 
\footnote{It is interesting to note that while the {\bf {field}} equations are always continuous in the charges, the {\bf spinor} constraints depend on charges discontiunously. The number of supersymmetry preserved depends on the number of non-zero charges, but not on the magnitudes of the charges.}.
When four charges are non-zero, one has to fix $\eta_*$ for consistency of the field equations, thereby set half of the spinor degree of freedom to zero. Supersymmetry is thus reduced by half. Each of the two spinor constraints in (\ref{NSSC}) reduces the (remaining) spinor degree of freedom by half. Thus only ${1 \over 8}$ (= ${1 \over 2^3}$) of the original $N=8$ supersymmetry is preserved. Same spinor constraints are obtained when only three charges are non-zero. Therefore the BPS-saturated states with four or three charges in the pure NS-NS sector preserve $N=1$ supersymmetry
\footnote{In the heterotic case [22], configurations with three to four non-zero charges also prserve $N=1$ supersymmetry, although the heterotic string only has $N=4$ supersymmetry, which is just half of that of the IIA superstring in 4d. The difference between the two cases is indicated by the present of $\eta_*$ in the IIA case. If $\eta_*$ has to be fixed in order to get the physical mass, supersymmetry is reduced from $N=8$ to $N=4$. After that, the patterns of supersymmetry breaking for the two cases are essentially the same.}.

Consider the case with two charges only. If both charges are of the same type, ${\it i.e.}$, both are electric or both are magnetic, $\eta_*$ has to be fixed. But only one of the two constraints in (\ref{NSSC}) remains. Therefore $N=2 (= 8 \times {1 \over 2^2})$ supersymmetry is preserved. When the two charges are of different types, ${\it e.g.}$, only $P^{NK}_1$ and $Q^{NK}_2$ are non-zero, both constraints in (\ref{NSSC}) are present, but $\eta_*$ does not need to be fixed. The configuration obtained by setting $P^{NF}_1 = Q^{NF}_2 = 0$ in (\ref{HETSOL}) can be considered as the solution of the KSEs with $\eta_* = 1, {\it i.e.}, (\varepsilon_u, \varepsilon_l) = (\varepsilon^+_l, \varepsilon^-_u)$ as well as the solution of KSEs with $\eta_* = -1, {\it i.e.}, (\varepsilon_u, \varepsilon_l) = (\varepsilon^+_u, \varepsilon^-_l)$. When the configuration is considered as the solution of the KSEs with either $\eta_* = 1$ or $\eta_* = -1$ exclusively, we see that it preserves $N=1 (= 8 \times {1 \over 2} \times {1 \over 2^2})$ supersymmetry. Therefore the supersymmetry transformations of the gravitinos and modulinos in (\ref{STG411}) vanish for the two different choices of sets of Killing spinors corresponding to $\eta_* = +1$ and -1 with the bosonic background defined in (\ref{HETSOL}). Therefore the configuration preserves $N=2 (= 1 + 1)$ supersymmetry
\footnote{If the non-zero charges include $P^{NF}_1$ or $Q^{NF}_2$, we can redefine $\eta_p$ or $\eta_q$, respectively, to remove the dependence on $\eta$ of the configuration.}.

When only one charge is non-zero, only one of the two constraints in (\ref{NSSC}) remains, and $\eta_*$ need not be fixed. So the solution preserves $N=4$ ( = 8 $ \times {1 \over 2} $ ) supersymmetry.

In summary, we conclude that the specification of each of the $\eta$s breaks ${1 \over 2}$ of the (remaining) supersymmetry. In the case of only one non-zero charge, ${1 \over 2}$ of the supersymmetry is broken, as only $\eta_p$ or $\eta_q$ need to be fixed (to maximize the mass), thus $N=4$ is preserved. In the case of two non-zero charges, only ${1 \over 2^2}$ of supersymmetry is preserved, as we need to fix two $\eta$s, ${\it i.e.}$, $\eta_*$ and $\eta_p (\eta_q)$ if both charges are magnetic (electric), $\eta_p$ and $\eta_q$ if the charges are of different types. Therefore $N=2$ supersymmetry is preserved. With three to four non-zero charges, all three $\eta$s need to be fixed, and only ${1 \over 2^3}$ of supersymmetry is preserved, ${\it i.e.}$, $N=1$. 

At this point, we try to find configurations that preserve $3 \over 8$ of the $N=8$ supersymmetry. From supersymmetry algebra, one may conclude that if there exists p different combinations of $\eta$s that give the same physical mass, which implies that there are p central charges coincide, then the corresponding configuration preserves $N=p$ supersymmetry. That is indeed true for the cases of $p=4, 2, 1$. However, it is not true when $p=3$. As an example, consider a configuration with the charges: $ (P^{NK}_1, P^{NF}_1, Q^{NK}_2, Q^{NF}_2) = (P, -P, Q, P)$, where $Q > P > 0$. Such charge assignment satisifies the inequalities: $P + Q = Q_R + P_R = Q_R - P_R = Q_L + P_L > ( Q_L - P_L ) > 0$, and leads to singular configurations. Correspondingly, three central charges coincide and are equal to the physical mass, $M = P + Q$. The three combinations of $\eta$s that give the same physical mass are $( \eta_*, \eta_q, \eta_p)$ = (1, 1, 1), (1, 1, -1), and (-1, 1, 1), from (\ref{NSMASS}). Although these three sets of $\eta$ combinations give the same space-time metric from (\ref{HETSOL}), they have different internal metric fields, $e_1^{\hat 1}$ and $e_2^{\hat 2}$. Therefore they correspond to different configurations, even though they correspond to the same mass. After checking out all cases,  we conclude that there is no solution (static, spherically symmetric, with no axion and off diagonal internal metric elements) that preserves $3 \over 8$ of the $N=8$ supersymmetry
\footnote{Similar situation is found in the case of heterotic string. When the central charges vanish, we get massless black hole solutions [16]. We found two different singular configurations corresponding to the two different choices of $(\eta_p, \eta_q)$, ${\it i.e.}$, $\eta_q = \eta_p$ or $\eta_q = - \eta_p$. Each of the two configurations preserves $1 \over 4$ of the original $N=4$ supersymmetry instead of one configuration preserving $N=2$ supersymmetry.}.

\subsection{Ramond-Ramond sector}

In this section, we turn off all NS-NS charges, and solve the KSEs with non-zero RR charges. In principle, these solutions can be obtained by performing U-duality on solution with NS-NS charges only. However, by solving the KSEs explicitly, we can study the pattern of supersymmetry breaking associated with each of the non-zero RR charges. It also provides a way of verifying the D-brane [32]-[36] intersection rules [36][37], at least in our chosen examples.

From the KSEs, (\ref{NSRM}) to (\ref{NSRTH}), we find that the KSEs with $\varepsilon^+_u$ on the left hand side have the same form as those with $\varepsilon^-_u$ on the left if we turn off the RR vector charges, $Q^{RK}$ and $P^{RK}$. Therefore we first consider the case when all NS-NS charges as well as the charges from RR vector field vanish (non-zero charges from RR vector field are considered in the following paragraphs). The KSEs, (\ref{NSRM}) - (\ref{NSRTH}), reduces to

\begin{eqnarray}
R \sqrt{\lambda} ( - \ln e_m + {1 \over 2} \sigma )' \varepsilon_{ul} =
{1 \over 2} \left( \pm {\bf{P}}^{RF}_{mi} + i {\bf{Q}}^{mi}_{RF} \right) \gamma^{mi} \varepsilon_{lu}
\nonumber 
\end{eqnarray}

\begin{eqnarray}
R \sqrt{\lambda} \left( \ln {\lambda} + \sigma \right)' \varepsilon_{ul} =\pm {1 \over 2} {\bf P}^{RF}_{ij} \gamma^{ij} \varepsilon_{lu}
\nonumber
\end{eqnarray}

\begin{eqnarray}
R \sqrt{\lambda} \left( \ln {\lambda} - \sigma \right)' \varepsilon_{ul} = - {i \over 2} {\bf Q}^{RF}_{ij} \gamma^{ij} \varepsilon_{lu} ,
\label{RSE}
\end{eqnarray}
where $\varepsilon_{ul} = \varepsilon^{\pm}_{ul}$, and $e_i \equiv e_i^{\hat i}$. There are two sets of consistent KSEs. One of them relates the upper and lower components of the spinors, $\varepsilon^+_{ul}$, which originate from the same 10-d spinor with positive chirality. The other set of KSEs relates the upper and lower components of spinors, $\varepsilon^-_{ul}$, originate from another 10-d spinor with negative chirality. As these two sets of KSEs turn out to be identical (apart from the chiralities of the Killing spinors), we write both of them together with $\varepsilon_{ul} = \varepsilon^{\pm}_{ul}$. In [38], the $Z_2$ element of the U-duality group, which maps all the NS-NS gauge fields to RR gauge fields and vice versa for IIA superstring compactified on $T^4$, was shown explicitly. The NS-NS charges of the configuration (\ref{HETSOL}) considered in Section IIIA, ${\it i.e.}$, $P^{NF}_1, P^{NK}_1, Q^{NK}_2$ and $Q^{NF}_2$, are mapped by the $Z_2$ element to the RR charges : $P_{12}, P_{34}, Q_{23}, Q_{14}$ (up to signs)
\footnote{ Another choice for non-zero RR charge assignment considered in [26] is : $Q_{ij}, P_{ik}$, where $i \not = j \not = k$. So the corresponding field configuration can have only two non-zero independent charges. Other charges are related to these two by $SO(6)$. Here we only look for configurations with maximum number of charges, ${\it i.e.}$, four charges.}.With this charge assignment, the KSEs are explicitly solved and the solution is as follows:

\begin{eqnarray}
{ \lambda = { r^2 \over { \left[ ( r + \eta_{12} P_{12} ) ( r + \eta_{34} P_{34} ) ( r + \eta_{23} Q_{23} ) ( r + \eta_{14} Q_{14} ) \right]^{1 \over 2} }}}
\nonumber 
\end{eqnarray}

\begin{eqnarray}
{ R = \left[ ( r + \eta_{12} P_{12} ) ( r + \eta_{34} P_{34} ) ( r + \eta_{23} Q_{23} ) ( r + \eta_{14} Q_{14} ) \right]^{1 \over 2} }
\nonumber 
\end{eqnarray}

\begin{eqnarray}
{ e_1 = \left[ {  { (r + \eta_{12} P_{12}) (r + \eta_{23} Q_{23}) } \over { (r + \eta_{34} P_{34}) (r + \eta_{14} Q_{14}) } } \right]^{1 \over 4} }
\nonumber 
\end{eqnarray}

\begin{eqnarray}
{ e_2 = \left[ {  { (r + \eta_{12} P_{12}) (r + \eta_{14} Q_{14}) } \over { (r + \eta_{34} P_{34}) (r + \eta_{23} Q_{23}) } } \right]^{1 \over 4} }
\nonumber 
\end{eqnarray}

\begin{eqnarray}
{ e_3 = \left[ {  { (r + \eta_{34} P_{34}) (r + \eta_{23} Q_{23}) } \over { (r + \eta_{12} P_{12}) (r + \eta_{14} Q_{14}) } } \right]^{1 \over 4} }
\nonumber 
\end{eqnarray}

\begin{eqnarray}
{ e_4 = \left[ {  { (r + \eta_{34} P_{34}) (r + \eta_{14} Q_{14}) } \over { (r + \eta_{12} P_{12}) (r + \eta_{23} Q_{23}) } } \right]^{1 \over 4} }
\nonumber 
\end{eqnarray}

\begin{equation}
{ e_{5,6} = \left[ {  { (r + \eta_{23} Q_{23}) (r + \eta_{14} Q_{14}) } \over { (r + \eta_{12} P_{12}) (r + \eta_{34} P_{34}) } } \right]^{1 \over 4} } ,
\label{RSOL}
\end{equation}
where $\eta_{12} \eta_{34} = \eta_{23} \eta_{14}$, and $\sigma = {\rm ln}\, {\rm det}\, e_m$. The dilaton does not run $( {\it i.e.} \phi = 0)$ as it only couples to the NS-NS charges, as shown in the Lagrangian (\ref{LIIAE}). Again we set the horizon to be at the origin of the radial direction.

We can interprete the above configuration in terms of D-branes, which are the RR charge carriers [32]. The configuration (\ref{RSOL}) corresponds to the intersection of two D-2-branes and two D-4-branes (${\it i.e.}, 2 \perp 2 \perp 4 \perp 4$)
\footnote{The configuration can also be interpreted as the intersection of two 2-branes and two 5-branes (${\it i.e.}, 2 \perp 2 \perp 5 \perp 5$) in 11-d, i.e., as configuration from intersecting M-branes [39]-[43]. Each of the two M-5-branes is parallel to the 11th dimension, while the two M-2-branes are orthogonal to it.}. 
The two electric charges, $Q_{23}$ and $Q_{14}$,  are carried by two D-2-branes wrapping around the compactified toroidal directions (23) and (14), respectively [38]. The magnetic charges, $P_{12}$ and $P_{34}$, are carried by two D-4-branes wrapping around the toroidal directions (3456) and (1256), respectively. With these identification for the directions of the D-branes, we can verify the D-brane intersection rules [36]. The two D-2-branes intersect each other at a point (the origin), while the two D-4-branes intersect each other at a D-2-brane (${\it i.e.}$, (56)). Each of the D-2-branes intersect each of the D-4-branes on a D-1-brane, (${\it i.e.}$, (23) intersects (3456) and (1256) on (3) and (2) respectively, (14) intersects (3456) and (1256) on (4) and (1) respectively). The four D-branes intersect at a point (the origin).

Each non-zero charge, $P_{12}, P_{34}, Q_{41}, Q_{23}$, creates a spinor constraint,

\begin{equation}
\varepsilon_u = \eta_{12} \gamma^{12} \varepsilon_l, \ \ \ \
\varepsilon_u = \eta_{34} \gamma^{34} \varepsilon_l, \ \ \ \
\varepsilon_u = i \eta_{14} \gamma^{41} \varepsilon_l, \ \ \ \
\varepsilon_u = -i \eta_{23} \gamma^{23} \varepsilon_l ,
\label{RSC}
\end{equation}
respectively, with $\eta_{12} \eta_{34} = \eta_{23} \eta_{14}$. Again, it is the same as expected from (\ref{GSC}). There are only three independent constraints in (\ref{RSC}). 

The mass of the black hole is

\begin{equation}
M = {1 \over 4} \left( \eta_{12} P_{12} + \eta_{34} P_{34} + \eta_{23} Q_{23} + \eta_{14} Q_{14} \right) .
\label{RMASS}
\end{equation}
In addition to the relation: $\eta_{12} \eta_{34} = \eta_{23} \eta_{14}$, the $\eta$s are also required to be chosen in such a way that the right hand side is maximized, which is then equal to the physical mass. That guarantees no massless solutions. Like the NS-NS case, a solution with three $\eta$s and four charges can be singular with certain types of charge assignment $({\it e.g.,} P_{12}, P_{34}, Q_{23} \gg - Q_{14} > 0)$. For regular solutions $({\it e.g.,} P_{12}, P_{34}, Q_{23}, Q_{14} > 0)$, the mass formular can be replaced by

\begin{equation}
M = {1 \over 4} \left( | P_{12} | + | P_{34} | + | Q_{23} | + | Q_{14} | \right) .
\label{RMASS2}
\end{equation}

For configurations with three charges or fewer, the three $\eta$s are enough to make the corresponding mass proportional to the sum of the absolute values of the charges. They have zero entropies, with the horizons coincide with the singular surfaces.

We study the pattern of supersymmetry breaking with our assumptions of zero charges from NS-NS fields and the RR vector field in this paragraph. If three to four charges are non-zero, we have three independent spinor constraints from (\ref{RSC}). Each reduces the spinor degree of freedom by half. Consequently the solution preserves $N=1 (=8 \times {1 \over 2^3})$ supersymmetry. With only two non-zero charges, two of the spinor constraints from (\ref{RSC}) survive, resulting in a configuration with $N=2$ (= $8 \times {1 \over 2^2} $) supersymmetry. With just one non-zero charge, the configuration preserves $N=4$ (= $8 \times {1 \over 2}$) supersymmetry, as only one of the spinor constraints in (\ref{RSC}) survives.  

It is interesting to check whether BPS-saturated states obtained by solving KSEs in RR sector under our working assumptions (i.e., spherical symmetric, static, only have scalars fields from dilaton and the diagonal internal metric, and zero charges from RR vector) can have more than four charges without referring to U-duality. As each magnetic charge leads to constraint of the form, $\varepsilon_u = \eta_{ij} \gamma^{ij} \varepsilon_l$, while each electric charge creates constraint of the form, $\varepsilon_u = i \eta_{ij} \gamma^{ij} \varepsilon_l$, we cannot have electric and magnetic charges with the same indices. Suppose we add the fifth charge $P_{13}$. The two spinor constraints from $P_{13}$ and $P_{12}$ imply that ${\varepsilon}_u$ has to be an egienvector of $\gamma^{23}$. But $\gamma^{23}$ does not commute with $\gamma^{13}$ (from (\ref{RSC}), $\varepsilon_u$ has to be an eigenvector of $\gamma^{13}$), therefore it results in imcompatible spinor constraints. Finally, if we add $P_{56}$, no such imcompatibility occurs, but the resulting configuration would over-constrain the spinor, ${\it i.e.}$, it only has two spinor degree of freedom. That is because the spinor constraint associated with $P_{56}$ is not derivable from (\ref{RSC}), and so there are four independent spinor constraints, thereby reducing the supersymmetry by a factor of $1 \over 2^4$. Similarly it is also impossible to add a fifth electric charge to the solution (\ref{RSOL}) consistently. Therefore, the BPS-saturated black hole solutions can at most have four non-zero charges from the Ramond-Ramond three-form fields.

\subsubsection{T-dual configurations}

The above configuration (\ref{RSOL}), corresponding to the intersecting D-brane configuration: $2 \perp 2 \perp 4 \perp 4$, is T-dual to a more symmetrical configuration [37] corresponding to the intersection of one D-0-brane and three D-4-branes, ${\it i.e.}$, $0 \perp 4 \perp 4 \perp 4$. By performing T-duality transformations
\footnote{Not the general T-duality transformations. We only consider the particular element of the T-duality group that inverse the radius of the corresponding compactified toroidal dimension.} 
on the 2nd and 3rd toroidal directions, the D-branes : (23),(14),(3456),(1256), are mapped to the D-branes : (),(1234),(2456),(1356). They carry the charges: $Q^{RK}, P_{56}, P_{13}$, and $P_{24}$, respectively. Again, the intersection rule is clearly verified. Each of the D-4-brane intersects another D-4-brane on a D-2-brane (${\it i.e.}$, (1234) intersects (2456) at (24), (2456) intersects (1356) at (56), (1234) intersects (1356) at (13)). The three D-2-branes intersect at a point, which couples to the 0-brane. As the particular element of the T-duality group that relates the above two configurations does not map a configuration with a diagonal internal metric and zero axion to a configuration with non-diagonal internal metric and non-zero axion [44], the configuration corresponding to the D-brane intersection : $0 \perp 4 \perp 4 \perp 4$, should also be a solution of our KSEs.

We have explicitly solved the KSEs with the charge assignment: $( Q^{RK}, P_{56}, P_{13}, P_{24} )$. The space-time metric, $\lambda$, has the expected form:

\begin{equation}
{ \lambda = { r^2 \over { \left[ ( r + \eta_{13} P_{13} ) ( r + \eta_{24} P_{24} ) ( r + \eta_{56} P_{56} ) ( r + \eta_q Q^{RK} ) \right]^{1 \over 2} }}}
\label{MIX1L}
\end{equation}
where $\eta_q$ is fixed by the spinor constraints. They are : 

\begin{equation}
\varepsilon_u = \eta_1 \gamma^{13} \varepsilon_l, \ \ \ \ \
\varepsilon_u = \eta_2 \gamma^{24} \varepsilon_l, \ \ \ \ \
\varepsilon_u = \eta_5 \gamma^{56} \varepsilon_l, \ \ \ \ \
\varepsilon_u = i \eta_e \eta_q \varepsilon_l .
\label{R2SC}
\end{equation}
where $\eta_e = \pm 1$ for $\varepsilon = \varepsilon^{\pm}$. These constraints agree with (\ref{GSC}). These constraints are not independent. They associate with the non-zero charges, $P_{13}, P_{24}, P_{56}$ and $Q^{RK}$, respectively. Using the fact that $ \gamma^{13} \gamma^{24} \gamma^{56} = - \gamma^7$, and the relations given in Section IIIB : $ \gamma^7 \varepsilon^{\pm} = \pm i \varepsilon^{\pm}$, we find $\eta_q = \eta_1 \eta_2 \eta_5$, and hence there are only three independent spinor constraints in (\ref{R2SC}). Therefore the above solution contains three independent $\eta$s, and preserves $N=1 (=8 \times {1 \over 2^3})$ supersymmetry.

An observation about the electric charge from RR vector field is made. From KSEs view point, it is the only charge that can couple to the three RR magnetic charges, $P_{13}, P_{24}, P_{56}$, in a supersymmetric black hole solution. Any additional RR charges originate from the RR three-form fields would over-constrain the spinor, as has been discussed previously. The magnetic RR vector charge, $P^{RK}$, can not replace $Q^{RK}$ consistently because the associated spinor constraint is not consistent with those in (\ref{R2SC}). On the other hand, from the D-brane view point, only the electric D-0-brane can couple consistently with the three intersecting magnetic D-4-branes. Each pair of the D-4-branes intersect at a D-2-brane, and the resulting three D-2-branes can only intersect at a point which can only couple to a D-0-brane. The D-0-brane can only carry the electric charge from RR vector, as it has no index. Therefore the KSEs' method provides a consistency check of the intersection rule of D-branes in this case.

There is yet another configuration related to the above two configurations by T-duality. By doing T-duality on the 5th and 6th toroidal directions, the configuration: (),(1234),(2456),(1356), is tranformed to: (56),(123456),(24),(13), which carry the charges: $Q_{56}, P^{RK}, Q_{24},$ and $Q_{13}$, respectively. This configuration corresponds to an intersecting D-brane configuration in which three intersecting D-2-branes are all contained in a D-6-brane.

The spinor constraints of the configuration associated with $Q_{13}, Q_{24}, Q_{56}, P^{RK}$, respectively, are the following:

\begin{equation}
\varepsilon_u = i \eta_{13} \gamma^{13} \varepsilon_l, \ \ \ \ \
\varepsilon_u = i \eta_{24} \gamma^{24} \varepsilon_l, \ \ \ \ \
\varepsilon_u = i \eta_{56} \gamma^{56} \varepsilon_l, \ \ \ \ \
\varepsilon_u = - \eta_e \eta_p \varepsilon_l .
\label{R3SC}
\end{equation}
where $\eta_e = \pm 1$ for $\varepsilon = \varepsilon^{\pm}$. These constranits agree with (\ref{GSC}). Like the previous case, only three independent spinor constraints are in (\ref{R3SC}), and $\eta_p$ = $\eta_1 \eta_2 \eta_5$. The configuration preserves $N=1 (=8 \times {1 \over 2^3})$ supersymmetry.

\subsection{Configurations with both NS-NS and RR charges}

As each NS-NS charge relates the the two-component spinors $\varepsilon^{\pm}_{ul}$ to $\varepsilon^{\mp}_{lu}$, and each RR charge relates $\varepsilon^{\pm}_{ul}$ to $\varepsilon^{\pm}_{lu}$, therefore a configuration with both NS-NS charge(s) and RR charge(s) necessarily involve all the four types of two-component spinors, ${\it i.e.}$, $\varepsilon^+_u, \varepsilon^+_l, \varepsilon^-_u$, and $\varepsilon^-_l$.

We can consider the relations between the two-component spinors associated with a configuration with both NS-NS charges and RR charges as follows. Starting with $\varepsilon^+_u$, the NS-NS charges relate them with $\varepsilon^-_l$ by the associated spinor constraints from (\ref{GSC}), and the RR charges relate $\varepsilon^+_u$ to $\varepsilon^+_l$ with the appropriate spinor constraints from (\ref{GSC}). In the same way, we can start with $\varepsilon^-_u$ and put down their relations with $\varepsilon^+_l$ and $\varepsilon^-_l$ from the spinor constraints associated with the NS-NS charges and RR charges. These two sets of relations on the spinors, one starts with $\varepsilon^+_u$, the other starts with $\varepsilon^-_u$, has to be consistent. We thus get a necessary condition of getting consistent KSEs involving both NS-NS charges and RR charges.

We are going to find a configuration with $P^{RF}_{12}, Q^{RF}_{23}, P^{NK}_{1},$ and $Q^{NK}_{3}$. Unlike the previous two Sections, we discuss the spinor constraints in detail before giving the field configurations in order to illustrate the above mentioned criteria for getting consistent KSEs when both NS-NS charge(s) and RR charge(s) are non-zero.

Starting with $\varepsilon^+_u$, we get the following spinor constraints from (\ref{GSC}) associated with the charges: $P^{RF}_{12}, Q^{RF}_{23}, P^{NK}_{1},$ and $Q^{NK}_{3}$, respectively:

\begin{equation}
\varepsilon^+_u = \eta_{12} \gamma^{12} \varepsilon^+_l,\ \ \ \ \
\varepsilon^+_u = -i \eta_{23} \gamma^{23} \varepsilon^+_l, \ \ \ \ \
\varepsilon^+_u = i \eta_{1} \gamma^{1} \varepsilon^-_l, \ \ \ \ \
\varepsilon^+_u = - \eta_{3} \gamma^{3} \varepsilon^-_l ,
\label{M1SCA}
\end{equation}
with the relation: $\eta_1 \eta_3 = \eta_{12} \eta_{23}$. These four spinor constraints are not independent. One of the two constraints from the NS-NS (RR) charges can be derived from the two constraints from the RR (NS-NS) charges.

We can also start with $\varepsilon^-_u$, and obtain the following spinor constraints from (\ref{GSC}):

\begin{equation}
\varepsilon^-_u = \eta_{12} \gamma^{12} \varepsilon^-_l,\ \ \ \ \
\varepsilon^-_u = -i \eta_{23} \gamma^{23} \varepsilon^-_l, \ \ \ \ \
\varepsilon^-_u = i \eta_{1} \gamma^{1} \varepsilon^+_l, \ \ \ \ \
\varepsilon^-_u = - \eta_{3} \gamma^{3} \varepsilon^+_l .
\label{M1SCB}
\end{equation}
These two sets of spinor constraints, (\ref{M1SCA}) and (\ref{M1SCB}), are consistent with each other. Therefore, we can just consider that the spinors, $\varepsilon^+_u, \varepsilon^+_l, \varepsilon^-_l$ are related by (\ref{M1SCA}), and $\varepsilon^-_u$ is given by the first relation in (\ref{M1SCB}) for consistency. Note that the spinor constraints, (\ref{M1SCA}) and (\ref{M1SCB}), are the same as what we find after solving the KSEs explicitly.

With only non-zero NK fields and RF fields, the KSEs, (\ref{NSRM}) to (\ref{NSRTH}), are reduced to the following form:

\begin{equation}
R \sqrt{\lambda} ( - \ln e_m + \phi + {1 \over 2} \sigma )' \varepsilon^{\pm}_{ul} =
{1 \over 2} \left[ \left( \mp {\bf{Q}}^{NK}_m - i {\bf{P}}^m_{NK} \right) \gamma^m \varepsilon^{\mp}_{lu} + \left( \pm {\bf{P}}^{RF}_{mi} + i {\bf{Q}}^{mi}_{RF} \right) \gamma^{mi} \varepsilon^{\pm}_{lu}  \right] 
\label{MIX1A}
\end{equation}

\begin{equation}
R \sqrt{\lambda} ( \phi + {1 \over 2} \sigma )' \varepsilon^{\pm}_{ul} =
{1 \over 8} \left( \pm {\bf{P}}^{RF}_{ij} + i {\bf{Q}}_{ij}^{RF} \right) \gamma^{ij} \varepsilon^{\pm}_{lu}
\label{MIX1B}
\end{equation}

\begin{equation}
R \sqrt{\lambda} \left( \ln {\lambda} - \sigma \right) ' \varepsilon^{\pm}_{ul} = \mp {\bf Q}^{NK}_i \gamma^i \varepsilon^{\mp}_{lu} - {i \over 2} {\bf Q}_{ij}^{RF} \gamma^{ij} \varepsilon^{\pm}_{lu}  
\label{MIX1C}
\end{equation}

\begin{equation}
R \sqrt{\lambda} \left( \ln {\lambda} + \sigma \right) ' \varepsilon^{\pm}_{ul} = i {\bf P}^{NK}_i \gamma^i \varepsilon^{\mp}_{lu} \pm {1 \over 2} {\bf P}_{ij}^{RF} \gamma^{ij} \varepsilon^{\pm}_{lu} ,  
\label{MIX1D}
\end{equation}
The fact that the KSEs with $\varepsilon^+_{ul}$ on the left of the equalities have identical forms with those KSEs with $\varepsilon^-_{ul}$ on the left hand side is the reason that we choose the charges with non-zero NK fields and RF fields at the beginning.

Solving the KSEs, (\ref{MIX1A}) to (\ref{MIX1D}), with the charges: $P^{NK}_1, Q^{NK}_3, P^{RF}_{12}$, and $Q^{RF}_{23}$, we get the following solution:

\begin{eqnarray}
\lambda = { r^2 \over { \left[ ( r + \eta_1 P_{1} ) ( r + \eta_{12} P_{12} ) ( r + \eta_3 Q_{3} ) ( r + \eta_{23} Q_{23} ) \right]^{1 \over 2} }}   
\nonumber 
\end{eqnarray}

\begin{eqnarray}
R = \left[ ( r + \eta_1 P_{1} ) ( r + \eta_{12} P_{12} ) ( r + \eta_3 Q_{3} ) ( r + \eta_{23} Q_{23} ) \right]^{1 \over 2} 
\nonumber 
\end{eqnarray}

\begin{eqnarray}
e^{2 \phi} = \left[ { { r + \eta_1 P_1 } \over {r + \eta_3 Q_3} } \right]^{1 \over 2} 
\nonumber 
\end{eqnarray}

\begin{eqnarray}
e_1 = \left[ {  { (r + \eta_{12} P_{12}) (r + \eta_{23} Q_{23}) } \over { (r + \eta_1 P_{1})^2 } } \right]^{1 \over 4} 
\nonumber 
\end{eqnarray}

\begin{eqnarray}
e_2 = \left[ {  { r + \eta_{12} P_{12} } \over { r + \eta_{23} Q_{23}  } } \right]^{1 \over 4}    
\nonumber 
\end{eqnarray}

\begin{eqnarray}
e_3 = \left[ {  {  (r + \eta_3 Q_{3})^2 } \over { (r + \eta_{12} P_{12}) (r + \eta_{23} Q_{23}) } } \right]^{1 \over 4} 
\nonumber 
\end{eqnarray}

\begin{equation}
e_{4,5,6} = \left[ {  { r + \eta_{23} Q_{23} } \over { r + \eta_{12} P_{12}  } } \right]^{1 \over 4}  ,
\label{MIX1SOL} 
\end{equation}
where $\eta_1 \eta_3 = \eta_{12} \eta_{23}$. 

Like the configurations discussed in the previous two sections, the above configuration has a 10-d interpretation. It corresponds to a bound state between a magentic monople with charge $P^{NK}_1$, a D-2-brane (with charge $Q^{RF}_{23}$), a D-4-brane (with charge $P^{RF}_{12}$), and a fundamental string (with charge $Q^{NK}_3$) which lies on the intersection of the two D-branes and carries a momentum. Upon compactification, the D-2-brane wraps around the toroidal directions, (23), while the D-4-brane wraps on (3456). Bound states of fundamental string and D-branes have been studied in [45][46]. 

The mass of the black hole is

\begin{equation}
M = {1 \over 4} \left( \eta_1 P_{1} + \eta_3 Q_{3} + \eta_{12} Q_{12} + \eta_{23} Q_{23} \right) ,
\label{MIX1MA}
\end{equation}
where $\eta_1 \eta_3 = \eta_{12} \eta_{23}$. Like previous cases, the $\eta$s are chosen to maximize the above quantity, which would then be the physical mass. A configuration with four non-zero charges can be either singular (${\it e.g.}, P_1, Q_{23}, P_{12} \gg -Q_3 > 0$) or regular (${\it e.g.}, P_1, P_{12}, Q_3, Q_{23} > 0$), depending on the charges. Mass formular of the regular solution is 

\begin{equation}
M = {1 \over 4} \left( | P_{1} | + | P_{12} | + | Q_{3} | + | Q_{23} | \right).
\label{MIX1MB}
\end{equation}
For a configuration with three charges or less, the three $\eta$s are chosen such that the mass is equal to the sum of the absolute values of the charges. The corresponding black hole has zero entropy, and its horizon coincides with the singular surface.

Now we consider the pattern of supersymmetry breaking. For convenience, we choose to consider the constraints in (\ref{M1SCA}), which relate $\varepsilon^+_u$ with $\varepsilon^-_l$ and $\varepsilon^+_l$, and the first constraint in (\ref{M1SCB}), which relate $\varepsilon^-_u$ with $\varepsilon^-_l$. The two RR constraints determine $\varepsilon^+_l$ from $\varepsilon^+_u$, and force $\varepsilon^+_u$ to be an eigenvector of $\gamma^{13}$. Therefore, there are only four spinor degree of freedom left from the 16 spinor degree of freedom contained in $\varepsilon^+_l$ and $\varepsilon^+_u$. The two NS-NS constraints determine $\varepsilon^-_l$ from $\varepsilon^+_u$, and require $\varepsilon^+_u$ to be an eigenvector of $\gamma^{13}$, again. Therefore, the four spinor constraints in (\ref{M1SCA}) contain one redundant constraints, and imply a relation between the $\eta$s, ${\it i.e.}, \eta_1 \eta_3 = \eta_{12} \eta_{23}$. No spinor degree of freedom are available from $\varepsilon^-_u$, as it is fixed by the first constraint in (\ref{M1SCB}) for consistency. Therefore, out of the 32 spinor degree of freedom contained in $\varepsilon^+_u, \varepsilon^+_l, \varepsilon^-_l$ and $\varepsilon^-_u$, only four is unconstrained. Hence the configuration preserves $N=1$ supersymmetry. 

We can also find the number of supersymmetry preserved by considering the number of independent $\eta$s, like the previous Sections. After all, fixing an $\eta$ implies half of the (remaining) supersymmetry is broken by a certain spinor constraint described in (\ref{M1SCA}). Therefore, a configuration with three (two RR charges and one NS-NS charges, or vice versa) or four charges (two from each sector) preserves $N=1$ (= 8 $ \times {1 \over 2^3}$) supersymmetry, as each of the three spinor constraints associated with the fixing of the three $\eta$s reduce half of the (remaining) spinor degree of freedom. A configuration with two non-zero charges has two spinor constraints, therefore preserves $N=2 (= 8 \times {1 \over 2^2})$ supersymmetry. A configuration with only one non-zero charge has one spinor constraint, therefore preserves $N=4 (= 8 \times {1 \over 2})$ supersymmetry.

From the previous two subsections, IIIA and IIIB, we conclude that there has no configuration with three or more NS-NS (RR) charges, while still contain at least one RR (NS-NS) charge. That is because three NS-NS (RR) charges already reduce the supersymmetry from $N=8$ to $N=1$. Further constraint from a RR (NS-NS) charge would over-constrain the spinor, as the spinor constraint from a NS-NS charge has the form : $\varepsilon_l = C_1 \gamma^i \varepsilon_u$, while that from RR charge is : $\varepsilon_l = C_2 \gamma^{ij} \varepsilon_u$ for some complex numbers $C_i$, and therefore a NS-NS constraint can never be derivable from pure RR constraints, and vice versa.

\subsubsection{T-dual configurations}

In this subsection, we study the two configurations which are T-dual to the configuration, (\ref{MIX1SOL}). Recall that the above solution, with charges: $P^{NK}_1, Q^{RF}_{23}, P^{RF}_{12}$ and $Q^{NK}_{3}$, corresponds to a bound state of a magnetic monopole, a D-2-brane wrapping around the toroidal dimensions (23), a D-4-brane wrapping around the toroidal dimensions (3456), and a fundamental string lying on the intersection of the two D-branes, ${\it i.e.}$, the 3rd toroidal dimension, and carries a momentum. To get the first T-dual configuration, we act on the first configuration, which we obtain by explicitly solving the KSEs, with two T-duality transformations on the 2nd and 3rd toroidal dimensions. The new configuration then correspond to the bound state of a magnetic monopole, a D-0-brane, a D-4-brane wrapping around (2456), and a fundamental string which carries no momentum but has a non-zero winding number on the 3rd toroidal dimension.  The D-4-brane, the string, and the toroidal direction associated with the gauge field of the monopole, are orthogonal to each other. The corresponding charge configuration is: $P^{NK}_1, Q^{RK}, P^{RF}_{13}$ and $Q^{NF}_3$. 

We solve the KSEs, (\ref{NSRM}) to (\ref{NSRTH}), with the above charges, ${\it i.e.}$, $P^{NK}_1, Q^{RK}, P^{RF}_{13}$ and $Q^{NF}_3$. The spinor constraints associated with these charges respectively are:

\begin{equation}
\varepsilon^+_u = i \eta_{1} \gamma^{1} \varepsilon^-_l,\ \ \ \ \
\varepsilon^+_u = i \eta_q \varepsilon^+_l, \ \ \ \ \
\varepsilon^+_u = \eta_{13} \gamma^{13} \varepsilon^+_l, \ \ \ \ \
\varepsilon^+_u = \eta_{3} \gamma^{3} \varepsilon^-_l ,
\label{M2SC}
\end{equation}
there are only three independent spinor constraints, and the $\eta$s satisfy: $\eta_1 \eta_3 = \eta_q \eta_{13}$. By including the constraint: $\varepsilon^-_u = i \eta_1 \gamma^1 \varepsilon^+_l$, we get a set of consistent spinor constraints relating all the four different types of two-component spinors. As before, the constraints (\ref{M2SC}), which we obtain by explicitly solving the KSEs, is the same as expected from (\ref{GSC}). The configuration preserves $N=1$ supersymmetry.

This second configuration, with charges $P^{NK}_1, Q^{RK}, P^{RF}_{13}$ and $Q^{NF}_3$, was studied in [47], where the Bekenstein-Hawking entropy of the corresponding black hole is shown to coincide with the degeneracy of the corresponding stringy states. 

We can get a third configuration which is T-dual to the above one by acting on the above configuration with T-duality transformation on the 1st and 3rd toroidal directions. The resulting configuration corresponds to the bound state of a solitonic 5-brane wrapping around (23456), a D-2-brane wrapping around (13), a D-6-brane wrapping on the full torus, $T^6$, and a fundamental string lying on the intersection of the solitonic 5-brane and the D-2-brane and carries a momentum. These constituents of the bound state carry the respective charges: $P^{NF}_1, Q^{RF}_{13}, P^{RK}$ and $Q^{NK}_3$. The spinor constraints with these charges are consistent. The charge constraints associated with the above charges respectively are:

\begin{equation}
\varepsilon^+_u = i \eta_{1} \gamma^{1} \varepsilon^-_l,\ \ \ \ \
\varepsilon^+_u = i \eta_{13} \gamma^{13} \varepsilon^+_l, \ \ \ \ \
\varepsilon^+_u = - \eta_{p} \varepsilon^+_l, \ \ \ \ \
\varepsilon^+_u = \eta_{3} \gamma^{3} \varepsilon^-_l ,
\label{M3SC}
\end{equation}
where $\eta_p \eta_{13} = \eta_1 \eta_3$, and there are only three independent spinor constraints among the four in (\ref{M3SC}). For the consistency of the KSEs, we need the additional constraint on $\varepsilon^-_u$: $\varepsilon^-_u = \eta_p \varepsilon^-_l$. This configuration preserves $N=1$ supersymmetry. It has been studied in [40] and [48].

We would like to conclude this section by recalling from Section IIIC that one interesting implication of the spinor constraints stated in (\ref{GSC}) is that there is no state with both $P^{RK}$ and $P^{NK}$, or both $P^{RK}$ and $Q^{NF}$, under our assumptions of spherical symmetry, time-independence, and only the dilaton and the diagonal internal metric elements are non-zero among the scalar fields. In terms of bound states of objects in string theory, that means  that there is no bound state between a D-6-brane and a monopole, and also no bound state between a D-6-brane and a fundamental string with non-zero winding number, under our assumptions.

\section{Conclusion}

In this paper, we found a class of BPS-saturated black hole solutions of the low enery effective supergravity Lagrangian of toroidally compactified IIA superstring in four dimensions. We solved the Killing spinor equations (KSEs) explicitly under the assumptions of time-independence, spherical symmetry, and have turned off all the scalar fields except the dilaton and the diagonal internal metric elements. We find a set of spinor constraints associated with the different types of charges. In all the configurations considered in this paper, these rules were explicitly obtained by solving the corresponding KSEs. They governed the patterns of supersymmetry breaking.

The solutions in general could carry no more than four charges under the assumptions stated above
\footnote{We expect to find solutions with five independent charges when our assumptions are relaxed [31]. The most general dyonic BPS-saturated black hole solutions are expected to be obtained by performing U-duality on the solutions with five independent charges, ${\it i.e.}$, they are the generating solutions. It is in complete analogy with the case of heterotic string [24].}.
Configurations with three to four non-zero charges preserved $N=1$ supersymmetry. Those with two charges preserved $N=2$ supersymmetry, and configurations with one charges only preserved $N=4$ supersymmetry. We found no solutions that preserved $3 \over 8$ of the $N=8$ supersymmetry. Configurations with four non-zero charges might or might not be singular, depending on the signature of the charges. The mass of a black hole with no naked singularity was proportional to the sum of the absolute values of the charges, and it had non-zero entropy. Solutions with fewer than four charges had zero entropies, with the horizons coincided with the singular surfaces. Their masses were also proportional to the sum of the absolute values of their charges. 

There are three different types of solutions differentiated by the origin of their charges. They are the configurations with Neveu-Schwarz-Neveu-Schwarz charges only, configurations with Ramond-Ramond charges only, and the configurations with both Neveu-Schwarz-Neveu-Schwarz charges and Ramond-Ramond charges. 

In the first case, we solved the KSEs with non-zero Neveu-Schwarz-Neveu-Schwarz charges only. The KSEs have the same structure of the KSEs of the toroidally compactified heterotic string [22]. The KSEs for IIA superstring related the upper (lower) components of the spinors, which originate from a ten dimensional spinor with a particular chirality, to the lower (upper) components of other spinors, which originate from the other ten dimensional spinor with opposite chirality. We explicitly gave the map that related the spinors and charges from the compactified IIA superstring to that of the compactified heterotic string. With this map, we can reproduce the KSEs of the heterotic string from that of the IIA superstring.

In the second case, we solved the KSEs with Ramond-Ramond charges only. The configuration, which was U-dual to the Neveu-Schwarz-Neveu-Schwarz configuration found previously, was explicitly obtained. It corresponded to the intersecting D-brane configuration with two D-2-branes and two D-4-branes. The T-dual configurations, one corresponded to the intersection of a D-0-brane and three D-4-branes, and one corresponded to a D-6-brane containing three intersecting D-2-branes, were also shown to be solutions of the KSEs. We studied the corresponding pattern of supersymmetry breaking. The intersection rules [36] of the D-branes, which were defined in terms of individual D-brane, each of which carried only one unit of charge, were verified with the classical configurations which may contain very large charges.

In the final case, we solved KSEs with both Neveu-Schwarz-Neveu-Schwarz charges and Ramond-Ramond charges. We found three different solutions. Each contains two NS-NS charges and two RR charges, and is related to the others by T-duality. The first solution corresponded to a bound state of a monopole, a D-2-brane, a D-4-brane and a fundamental string which lies on the intersection of the two D-branes, and carries a momentum. The second corresponded to a bound state of a monopole, a D-0-brane, a D-4-brane, and a fundamental string with non-zero winding number [47]. The D-4-brane, the fundamental string, and the toroidal direction associated with the gauge field that supported the charge of the magnetic monopole, are orthogonal to each other. The third configuration corresponded to a bound state of the solitonic 5-brane, a D-6-brane, a D-2-brane, and a fundamental string which lies on the intersection of the D-2-brane and the solitonic 5-brane and carries a momentum [40][48]. 

\section{Acknowledgments}
I am grateful to Mirjam Cveti\v c for many useful and enlightening discussions in the whole project. This work could never have been completed without her continuous advices. The work was supported in part by U.S. Department of Energy Grant No. DOE-EY-76-02-3071 and the National Science Foundation PHY95-12732.

\section{Appendix}

In this Appendix, we consider the Majorana condition and the Weyl condition on a spinor, $\varepsilon_d$ ($d=10, 11$), in ten dimensions and in eleven dimensions. The $d$ dimensional Lorentz group is considered as a direct product of the four-dimensional Lorentz group and the six or seven dimensional rotation group in 10d and 11d, respectively.

When the d-dimensional Lorentz group is decomposed as follows :

\begin{equation}
SO(1,d-1) \rightarrow SO(1,3) \otimes SO(d-4) ,
\label{SORED}
\end{equation}
the corresponding $\gamma$ matrixes can be taken as

\begin{equation}
\Gamma^{\alpha} = \gamma^{\alpha} \otimes I , \ \ \ \ \ \ 
\Gamma^a = \gamma^5 \otimes \gamma^a , 
\label{GAMRED}
\end{equation}
where $\alpha$ = 0, 1,  2, 3, and $a$ = 1,...,d-4, as defined previously in Section III. The spinor can then be labelled by two indices, $\varepsilon^{{\bf a, m}}_d$, where ${\bf a}$ is the index for the spinor representation of $SO(1,3)$ and ${\bf m}$ is the index for the spinor representation of $SO(d-4)$. 

A d-dimensional spinor, $\varepsilon_d$, is Majorana if it satisfies :

\begin{equation}
{\bar \varepsilon_d} = \varepsilon_d^T \ \ C_d ,
\label{MAJ}
\end{equation}
where

\begin{equation}
{\bar \varepsilon_d} \equiv \varepsilon_d^{\dag} \Gamma^0 ,
\label{BAR}
\end{equation}
and the d-dimensional charge conjuation matrix, $C_d$, satisfies

\begin{equation}
C_d \Gamma^{\mu} C_d^{-1} = - \Gamma^{{\mu} T} , \ \ \ \ \ 
\mu = 0,1,...,(d-1) .
\label{CCM}
\end{equation}

Under the reduction (\ref{SORED}), and with the representation chosen for the $\Gamma$ matrixes (\ref{GAMRED}), $C_d$ can be written as a direct product:

\begin{equation}
C_d = C_4 \otimes C_{d-4} .
\label{CCC}
\end{equation}
However, as $\gamma^5$ satisfies the relation: 

\begin{equation}
\gamma^{5T} = C_4 \gamma^5 C_4^{-1} ,
\label{GA5}
\end{equation}
and $\gamma^a$ is antisymmetric for all $a = 1,2,...,(d-4)$, we find that $C_d$ given by (\ref{CCC}) with $C_{d-4} = I_{d-8}$ satisfies (\ref{CCM}). As $\Gamma^0 = \gamma^0 \otimes I_{d-4}$, we see that imposing the Majorana condition on the d-dimensional spinor, $\varepsilon^{{\bf a,m}}_d$, is equivalent of imposing the Majorana condition only on the spinor representation labelled by the index ${\bf a}$ of the 4-d Lorentz group, $SO(1,3)$. 

We now consider $\varepsilon_{11}$ and $\varepsilon_{10}$ seperately.  Only Majorana condition can be imposed on the 11-d spinor, $\varepsilon_{11}$. That is because Majorana condition and Weyl condition are imcompactible in 4-d and so we can only impose Majorana condition on the 4-d spinor representation labelled by ${\bf a}$. We cannot impose Weyl condition on the spinor representation labelled by ${\bf m}$, as no Weyl condition can be imposed on the spinor representation of $SO(7)$. Therefore the 11-d spinor can only be Majorana and have 32 real components. On the other hand, we can impose Majorana condition on $\varepsilon_{10}$ by imposing Majorana condition on the 4-d spinor representation labelled by ${\bf a}$, and also impose Weyl condition on the spinor representation of $SO(6)$ labelled by ${\bf m}$, by choosing the spinor to be an eigenvector of  $\gamma^7 (\equiv \gamma^1 \times \gamma^2 \times .... \times \gamma^6)$. Therefore the 10-d spinor can be both Majorana and Weyl and have 16 spinor degree of freedom.

\end{document}